
\def\unlock{
 \catcode`@=11 }% This allows us to modify PLAIN macros.
\unlock
\def\lock{\catcode`@=12 }% at signs are no longer letters

%%% special local internal skips
\newskip\@@@@@
\newskip\@@@@@two
\newskip\t@mp
%%%%%
%%%%%         FONT DEFINITIONS AND SETS
%%%%%
 
     %        \ten<>, \seven<>, \five<> defined in PLAIN TeX
     %            for rm, bf, i, sy
     %            all \ten<> are defined in PLAIN TeX
 
 \font\twentyfourrm=cmr10               scaled\magstep4
 \font\seventeenrm=cmr10                scaled\magstep3
 \font\fourteenrm=cmr10                 scaled\magstep2
 \font\twelverm=cmr10                   scaled\magstep1
 \font\ninerm=cmr9            \font\sixrm=cmr6
 \font\twentyfourbf=cmbx10              scaled\magstep4
 \font\seventeenbf=cmbx10               scaled\magstep3
 \font\fourteenbf=cmbx10                scaled\magstep2
 \font\twelvebf=cmbx10                  scaled\magstep1
 \font\ninebf=cmbx9            \font\sixbf=cmbx5
 \font\twentyfouri=cmmi10 scaled\magstep4  \skewchar\twentyfouri='177
 \font\seventeeni=cmmi10  scaled\magstep3  \skewchar\seventeeni='177
 \font\fourteeni=cmmi10   scaled\magstep2  \skewchar\fourteeni='177
 \font\twelvei=cmmi10     scaled\magstep1  \skewchar\twelvei='177
 \font\ninei=cmmi9                         \skewchar\ninei='177
 \font\sixi=cmmi6                          \skewchar\sixi='177
 \font\twentyfoursy=cmsy10 scaled\magstep4 \skewchar\twentyfoursy='60
 \font\seventeensy=cmsy10  scaled\magstep3 \skewchar\seventeensy='60
 \font\fourteensy=cmsy10   scaled\magstep2 \skewchar\fourteensy='60
 \font\twelvesy=cmsy10     scaled\magstep1 \skewchar\twelvesy='60
 \font\ninesy=cmsy9                        \skewchar\ninesy='60
 \font\sixsy=cmsy6                         \skewchar\sixsy='60
 \font\twentyfourex=cmex10  scaled\magstep4
 \font\seventeenex=cmex10   scaled\magstep3
 \font\fourteenex=cmex10    scaled\magstep2
 \font\twelveex=cmex10      scaled\magstep1
 \font\elevenex=cmex10      scaled\magstephalf
 \font\twentyfoursl=cmsl10  scaled\magstep4
 \font\seventeensl=cmsl10   scaled\magstep3
 \font\fourteensl=cmsl10    scaled\magstep2
 \font\twelvesl=cmsl10      scaled\magstep1
 \font\ninesl=cmsl9
 \font\twentyfourit=cmti10 scaled\magstep4
 
 \font\fourteenit=cmti10   scaled\magstep2
 \font\twelveit=cmti10     scaled\magstep1
 \font\twentyfourtt=cmtt10 scaled\magstep4
 \font\twelvett=cmtt10     scaled\magstep1
 \font\twentyfourcp=cmcsc10 scaled\magstep4
 \font\twelvecp=cmcsc10    scaled\magstep1
 \font\tencp=cmcsc10
 \font\hepbig=cmdunh10     scaled\magstep5
 \font\hep=cmdunh10        scaled\magstep0
 \newfam\cpfam
 \font\tenfib=cmr10  % quick fix for a missing font
 \newcount\f@ntkey            \f@ntkey=0
 \def\samef@nt{\relax \ifcase\f@ntkey \rm \or\oldstyle \or\or
          \or\it \or\sl \or\bf \or\tt \or\caps \fi }
 
 %%%%%
 %%%%%           FONT SETS
 %%%%%
 \def\twentyfourpoint{\relax
     \textfont0=\twentyfourrm           \scriptfont0=\seventeenrm
                                        \scriptscriptfont0=\fourteenrm
     \def\rm{\fam0 \twentyfourrm \f@ntkey=0 }\relax
     \textfont1=\twentyfouri            \scriptfont1=\seventeeni
                                        \scriptscriptfont1=\fourteeni
      \def\oldstyle{\fam1 \twentyfouri\f@ntkey=1 }\relax
     \textfont2=\twentyfoursy           \scriptfont2=\seventeensy
                                        \scriptscriptfont2=\fourteensy
     \textfont3=\twentyfourex           \scriptfont3=\seventeenex
                                        \scriptscriptfont3=\fourteenex
     \def\it{\fam\itfam \twentyfourit\f@ntkey=4 }\textfont\itfam=\twentyfourit
     \def\sl{\fam\slfam \twentyfoursl\f@ntkey=5 }\textfont\slfam=\twentyfoursl
                                        \scriptfont\slfam=\seventeensl
     \def\bf{\fam\bffam \twentyfourbf\f@ntkey=6 }\textfont\bffam=\twentyfourbf
                                        \scriptfont\bffam=\seventeenbf
                                        \scriptscriptfont\bffam=\fourteenbf
     \def\tt{\fam\ttfam \twentyfourtt \f@ntkey=7 }\textfont\ttfam=\twentyfourtt
     \def\caps{\fam\cpfam \twentyfourcp \f@ntkey=8 }
                                        \textfont\cpfam=\twentyfourcp
     \setbox\strutbox=\hbox{\vrule height 23pt depth 8pt width\z@}
     \samef@nt}
 \def\fourteenpoint{\relax
     \textfont0=\fourteenrm          \scriptfont0=\tenrm
     \scriptscriptfont0=\sevenrm
      \def\rm{\fam0 \fourteenrm \f@ntkey=0 }\relax
     \textfont1=\fourteeni           \scriptfont1=\teni
     \scriptscriptfont1=\seveni
      \def\oldstyle{\fam1 \fourteeni\f@ntkey=1 }\relax
     \textfont2=\fourteensy          \scriptfont2=\tensy
     \scriptscriptfont2=\sevensy
     \textfont3=\fourteenex          \scriptfont3=\twelveex
     \scriptscriptfont3=\tenex
     \def\it{\fam\itfam \fourteenit\f@ntkey=4 }\textfont\itfam=\fourteenit
     \def\sl{\fam\slfam \fourteensl\f@ntkey=5 }\textfont\slfam=\fourteensl
     \scriptfont\slfam=\tensl
     \def\bf{\fam\bffam \fourteenbf\f@ntkey=6 }\textfont\bffam=\fourteenbf
     \scriptfont\bffam=\tenbf     \scriptscriptfont\bffam=\sevenbf
     \def\tt{\fam\ttfam \twelvett \f@ntkey=7 }\textfont\ttfam=\twelvett
     \def\caps{\fam\cpfam \twelvecp \f@ntkey=8 }\textfont\cpfam=\twelvecp
     \setbox\strutbox=\hbox{\vrule height 12pt depth 5pt width\z@}
     \samef@nt}
 
 \def\twelvepoint{\relax
     \textfont0=\twelverm          \scriptfont0=\ninerm
     \scriptscriptfont0=\sixrm
      \def\rm{\fam0 \twelverm \f@ntkey=0 }\relax
     \textfont1=\twelvei           \scriptfont1=\ninei
     \scriptscriptfont1=\sixi
      \def\oldstyle{\fam1 \twelvei\f@ntkey=1 }\relax
     \textfont2=\twelvesy          \scriptfont2=\ninesy
     \scriptscriptfont2=\sixsy
     \textfont3=\twelveex          \scriptfont3=\elevenex
     \scriptscriptfont3=\tenex
     \def\it{\fam\itfam \twelveit \f@ntkey=4 }\textfont\itfam=\twelveit
     \def\sl{\fam\slfam \twelvesl \f@ntkey=5 }\textfont\slfam=\twelvesl
     \scriptfont\slfam=\ninesl
     \def\bf{\fam\bffam \twelvebf \f@ntkey=6 }\textfont\bffam=\twelvebf
     \scriptfont\bffam=\ninebf     \scriptscriptfont\bffam=\sixbf
     \def\tt{\fam\ttfam \twelvett \f@ntkey=7 }\textfont\ttfam=\twelvett
     \def\caps{\fam\cpfam \twelvecp \f@ntkey=8 }\textfont\cpfam=\twelvecp
     \setbox\strutbox=\hbox{\vrule height 10pt depth 4pt width\z@}
     \samef@nt}
 
 \def\tenpoint{\relax
     \textfont0=\tenrm          \scriptfont0=\sevenrm
     \scriptscriptfont0=\fiverm
     \def\rm{\fam0 \tenrm \f@ntkey=0 }\relax
     \textfont1=\teni           \scriptfont1=\seveni
     \scriptscriptfont1=\fivei
     \def\oldstyle{\fam1 \teni \f@ntkey=1 }\relax
     \textfont2=\tensy          \scriptfont2=\sevensy
     \scriptscriptfont2=\fivesy
     \textfont3=\tenex          \scriptfont3=\tenex
     \scriptscriptfont3=\tenex
     \def\it{\fam\itfam \tenit \f@ntkey=4 }\textfont\itfam=\tenit
     \def\sl{\fam\slfam \tensl \f@ntkey=5 }\textfont\slfam=\tensl
     \def\bf{\fam\bffam \tenbf \f@ntkey=6 }\textfont\bffam=\tenbf
     \scriptfont\bffam=\sevenbf     \scriptscriptfont\bffam=\fivebf
     \def\tt{\fam\ttfam \tentt \f@ntkey=7 }\textfont\ttfam=\tentt
     \def\caps{\fam\cpfam \tencp \f@ntkey=8 }\textfont\cpfam=\tencp
     \setbox\strutbox=\hbox{\vrule height 8.5pt depth 3.5pt width\z@}
     \samef@nt}
 
%%%%
%%%% Special notations and symbols
%%%%
 
   %%% Lagrangian symbol

%quick fix for a jam
 \newtoks\date
 \def\monthname{\relax\ifcase\month 0/\or January\or February\or
    March\or April\or May\or June\or July\or August\or September\or
    October\or November\or December\else\number\month/\fi}
 \date={\monthname\ \number\day, \number\year}
 \def\today{\the\day\ \monthname\ \the\year}
 \def\ie{\hbox{\it i.e.}}     
      
         \def\etal{\hbox{\it et.al.}}
 \def\\{\relax\ifmmode\backslash\else$\backslash$\fi}
 \def\nextline{\unskip\nobreak\hskip\parfillskip\break}
 
 \def\journal#1&#2,#3(#4){\unskip \enskip {\sl #1}~{\bf #2}, {\rm #3 (19#4)}}

 \def\topspace{\hrule height 0pt depth 0pt \vskip}
\def\nullbox#1#2#3{\vbox to #1
     {\vss\vtop to#2
       {\vss\hbox to #3 {}}}}
\def\n@ll{\nullbox{5pt}{3pt}{2pt}}    % NEED TO UNLOCK '@' to use
\def\UNRTXT#1{\vtop{\hbox{#1}\kern 1pt \hrule}}
\def\undertext#1{\ifvmode\ifinner \UNRTXT{#1}
                         \else    $\hbox{\UNRTXT{#1}}$
                         \fi
                 \else   \ifmmode \hbox{\UNRTXT{#1}}
                         \else    \UNRTXT{#1}
                         \fi
                 \fi }
 
%%%%%%%
%%%%%%%             HIGH ENERGY NOTATIONS
%%%%%%%
 
%%%%%
%%%%%   GAMMA MATRICIES
%%%%%
 
  %% subscripted notation
\def\gmu{\gamma^\mu } %% superscripted notation
 
%%%%%
%%%%%   BRANCHING RATIOS
%%%%%

%%%%%
%%%%%          Electron volt notations
%%%%%
 
\def\ev{{\rm e\kern-.1100em V}}
\def\tev{{\rm T\kern-.1000em \ev}}
\def\gev{{\rm G\kern-.1000em \ev}}
\def\mev{{\rm M\kern-.1000em \ev}}
\def\kev{{\rm K\kern-.1000em \ev}}
 
%%%%%
%%%%%           MASS SCALE NOTATION
%%%%%

%%%%%
%%%%%           PARTICLE NOTATIONS
%%%%%
 
\def\eright{e^-\kern-.3em\low{R}}      %% right handed electron
\def\eleft{e^-\kern-.3em\low{L}}       %% left handed electron
\def\nue{\nu\kern-.3em\low{e}{}}

%%%%%%%%
%%%%%%%%            MATHEMATIC NOTATION
%%%%%%%%
 
%%%
%%%           SUB/SUPER SCRIPTS
%%%
 
\def\low#1{\kern-0.11em\lower0.4em\hbox{$\scriptstyle #1 $}\hskip-0.08em}
\def\up#1{\kern-0.45em\raise0.4em\hbox{$\scriptstyle #1 $}\hskip0.29em}
 
%%%
%%%             OPERATIONS
%%%

 \let\int=\intop         
 \def\prop{\mathrel{{\mathchoice{\pr@p\scriptstyle}{\pr@p\scriptstyle}{
                 \pr@p\scriptscriptstyle}{\pr@p\scriptscriptstyle} }}}
 \def\pr@p#1{\setbox0=\hbox{$\cal #1 \char'103$}
    \hbox{$\cal #1 \char'117$\kern-.4\wd0\box0}}
 \let\sec@nt=\sec
 \def\sec{\relax\ifmmode\let\n@xt=\sec@nt\else\let\n@xt\section\fi\n@xt}
 \def\@versim#1#2{\lower0.2ex\vbox{\baselineskip\z@skip\lineskip\z@skip
   \lineskiplimit\z@\ialign{$\m@th#1\hfil##\hfil$\crcr#2\crcr\sim\crcr}}}
\def\lsim{\mathrel{\mathpalette\@versim<}}%Similar or less than%
\def\gsim{\mathrel{\mathpalette\@versim>}}%similar or greater than%

\def\barr#1{ \overline #1 }
   %%% bar superscript %%%
 
%%%%
%%%% DERIVATIVE OPERATORS
%%%%

\def\part#1#2{\partial^{#1}\kern-.1667em\hbox{$#2$}}
\def\der#1#2{d^{#1}\kern-.1067em\hbox{$#2$}}
          %%% four derivative super
       %%% four derivative sub

\def\delt#1{{\delta\kern-.15em #1 }}
 
%%%%
%%%%           FOUR VECTOR NOTATIONS.
%%%%
%%%%     NOTE:  FOUR DERIVATIVE OPERATORS ARE ABOVE
%%%%

%%%%
%%%%        OPERATORS
%%%%
 
%  Box operator
\def\sqr#1#2{{\vcenter{\hrule height.#2pt
    \hbox{\vrule width.#2pt height#1pt \kern#1pt
    \vrule width.#2pt}
    \hrule height.#2pt}}}

%%%%%
%%%%%       SPECIAL FUNCTIONS
%%%%%
 
  %% redefine the vector notation
                                        %%   arrow covers entire symbol
   %% forms the cross product
 %% forms the dot product

     %%% this yields the form A --> A' = A B C
     %%% where A(#1) is the function being
     %%% being transformed, B(#2) is the
     %%% operation in the transform,
     %%% C(#3) is the phase shift

     %%%  where #1 is the function name
     %%%        #2 is the function variables
     %%%        #3 is the phase shift
\def\gmumat#1#2{\barr{#1}\gmu \kern0.008in {#2}}
     %%%  where #1 is the antiparticle in the Lagrangian
     %%%        #2 is the particle in the Lagrangian

\def\coeff#1#2{{#1 \over #2 }}
   %% Builds fractions in math modes

   %% a group of indicies #1 assigned allowed values #2
   %% where #2 is in matrix notation
\def\rightbracearrow{\raise0.5em\hbox{$|$}\mkern-8.0mu\rightarrow}
   %% builds the construction |->    for nuclear decays

%%%%%%%
%%%%%%%          GROUP THEORY NOTATIONS
%%%%%%%
 
%%%%
%%%%             SYMMETRY GROUP NOTATION
%%%%
 
\def\U#1{U\kern-.25em \left( #1 \right) }
    % Normal group theory notation U(n)

    % Special group theory notation SU(N)
 
%%%%
%%%%           SYMMETRY FIELD NOTATION
%%%%
 
%% U(1) field notation
 
   %% superscripted notation
  %% subscripted notation
 
%% U(2) field notation
 
 %% subscripted notation
 %% superscripted notation
 
%% U(3) field notation
 
 %% subscripted notation
 %% superscripted notation
 
%% Color Space

%%%%%%
%%%%%%   These partial remains are from the phyzzx macro
%%%%%%
 
%%%%             LINE SPACING MACROS
%%%%     COPIED FROM THE SLAC PHYZZX MACRO PACKAGE
%%%%
 
 \normalbaselineskip = 20pt plus 0.2pt minus 0.1pt
 \normallineskip = 1.5pt plus 0.1pt minus 0.1pt
 \normallineskiplimit = 1.5pt
 \newskip\normaldisplayskip
    \normaldisplayskip = 20pt plus 5pt minus 10pt
 \newskip\normaldispshortskip
    \normaldispshortskip = 6pt plus 5pt
 \newskip\normalparskip
    \normalparskip = 6pt plus 2pt minus 1pt
 \newskip\skipregister
    \skipregister = 5pt plus 2pt minus 1.5pt
 \newif\ifsingl@    \newif\ifdoubl@
 \newif\ifb@@kspace   \b@@kspacefalse
 \newif\iftwelv@    \twelv@true
 \newif\ifp@genum
 
%%%%
%%%%        LINE SPACING CHOICES
%%%%
 
 \def\singlespace{\singl@true\doubl@false\b@@kspacefalse\spaces@t}
 \def\doublespace{\singl@false\doubl@true\b@@kspacefalse\spaces@t}
 \def\normalspace{\singl@false\doubl@false\b@@kspacefalse\spaces@t}
 \def\bookspace{\b@@kspacetrue\singl@false\doubl@false\spaces@t}
 \def\tablespace{\singlespace}
 \def\smashspace{\singl@false\doubl@false\b@@kspacefalse\subspaces@t3:8;}
 
%%%%
%%%%        FONT CHOICES
%%%%
 
 \def\Tenpoint{\tenpoint\twelv@false\spaces@t}
 \def\Twelvepoint{\twelvepoint\twelv@true\spaces@t}
 
%%%
%%%       SPACING CONTROLS
%%%
 
 \def\spaces@t{\relax
       \iftwelv@ \ifsingl@\subspaces@t3:4;\else\subspaces@t1:1;\fi
        \else \ifsingl@\subspaces@t1:2;\else\subspaces@t4:5;\fi \fi
                    %%%%%%original 3:5;
       \ifb@@kspace
          \baselineskip=14pt
       \fi
       \ifdoubl@ \multiply\baselineskip by 5
          \divide\baselineskip by 4 \fi }
 \def\subspaces@t#1:#2;{
       \baselineskip = \normalbaselineskip
       \multiply\baselineskip by #1 \divide\baselineskip by #2
       \lineskip = \normallineskip
       \multiply\lineskip by #1 \divide\lineskip by #2
       \lineskiplimit = \normallineskiplimit
       \multiply\lineskiplimit by #1 \divide\lineskiplimit by #2
       \parskip = \normalparskip
       \multiply\parskip by #1 \divide\parskip by #2
       \abovedisplayskip = \normaldisplayskip
       \multiply\abovedisplayskip by #1 \divide\abovedisplayskip by #2
       \belowdisplayskip = \abovedisplayskip
       \abovedisplayshortskip = \normaldispshortskip
       \multiply\abovedisplayshortskip by #1
         \divide\abovedisplayshortskip by #2
       \belowdisplayshortskip = \abovedisplayshortskip
       \advance\belowdisplayshortskip by \belowdisplayskip
       \divide\belowdisplayshortskip by 2
       \smallskipamount = \skipregister
       \multiply\smallskipamount by #1 \divide\smallskipamount by #2
       \medskipamount = \smallskipamount \multiply\medskipamount by 2
       \bigskipamount = \smallskipamount \multiply\bigskipamount by 4 }
 \def\normalbaselines{ \baselineskip=\normalbaselineskip
    \lineskip=\normallineskip \lineskiplimit=\normallineskip
    \iftwelv@\else \multiply\baselineskip by 4 \divide\baselineskip by 5
      \multiply\lineskiplimit by 4 \divide\lineskiplimit by 5
      \multiply\lineskip by 4 \divide\lineskip by 5 \fi }
 
%%%%%%
%%%%%%           CHAPTERS, SECTIONS, SUBSECTIONS
%%%%%%
%%%%%%           This section here the labeling
%%%%%%              of chapters, sections and subsections
%%%%%%           This section also builds the table of
%%%%%%              contents
%%%%%%
 
%%%%
%%%%                  PARAMETERS---LOTS OF THEM
%%%%
 
 \newskip\tablelineskip           \tablelineskip=0.7in
 \newskip\figurelineskip          \figurelineskip=1.55in
 \newskip\tablelinelength     \tablelinelength=4.5in
 \newskip\tabledotvskip       \tabledotvskip=-0.359in
 \newskip\tabledothskip       \tabledothskip=-0.306in%
 \newskip\bookchapterskip     \bookchapterskip=1.0in
 \newcount\chapternumber    \chapternumber=0
 \newcount\appendixnumber   \appendixnumber=0
 \newcount\sectionnumber    \sectionnumber=0
 \newcount\subsectionnumber \subsectionnumber=0
 \newcount\equanumber       \equanumber=1
 \newcount\problemnumber    \problemnumber=0
 \newcount\figurecount      \figurecount=1
 \newcount\conpage          \conpage=0
 \let\chapterlabel=0
 \newtoks\constyle          \constyle={\Number}
 \newtoks\appendixstyle     \global\appendixstyle={\Alphabetic}
 \newtoks\chapterstyle      \chapterstyle={\Number}
 \newtoks\subsecstyle       \subsecstyle={\alphabetic}
 \newskip\chapterskip       \chapterskip=\bigskipamount
 \newskip\sectionskip       \sectionskip=\medskipamount
 \newskip\headskip          \headskip=8pt plus 3pt minus 3pt
 \newif\ifsp@cecheck
 \newif\iffirst@ppendix       \global\first@ppendixtrue
 \newdimen\chapterminspace    \chapterminspace=15pc
 \newdimen\sectionminspace    \sectionminspace=8pc
 \interlinepenalty=50
 \interfootnotelinepenalty=5000
 \predisplaypenalty=9000
 \postdisplaypenalty=500
 \newwrite\tableconwrite
 \newbox\tableconbox
 \newif\iftabl@conlist
 \newwrite\figwrite
 \newwrite\figurewrite
 \newif\iffigur@list
 \newwrite\eqnwrite
 \newif\if@qlist
 \newif\ifeqlo@d            \eqlo@dfalse
 \newwrite\tablewrite
 \newwrite\tableswrite
 \newif\iftabl@list
 \newwrite\appendixwrite
 \newif\if@ppendix
 \newcount\@ppcharnumber     \@ppcharnumber=64
 \newcount\referencecount \newbox\referencebox
 \newcount\tablecount
 \newif\ifindex      \indexfalse%%% used in the index macros
 \newwrite\indexwrite
 \newbox\indexbox
 \newskip\indexskip
 \newif\ifmanualpageno   \manualpagenofalse
%    Stuff for Postscript file insertion
 \newcount\UGscale   \newdimen\UGdscale
 \newcount\UGleft    \newdimen\UGdleft
 \newcount\UGbot     \newdimen\UGdbot
 \newdimen\UGabs     \UGabs=.01in
 
%%%%
%%%%           LABLE STYLE CONTROLS
%%%%
 
 \def\Number#1{\number #1}
 \def\makel@bel{\xdef\chapterlabel{%
     \the\chapterstyle{\the\chapternumber}.}}
 \def\sectionlabel{\number\sectionnumber }
 \def\subseclabel{{\the\subsecstyle{\the\subsectionnumber}. }}
 \def\alphabetic#1{\count255='140 \advance\count255 by #1\char\count255}
 \def\Alphabetic#1{\count255=64 \advance\count255 by #1\char\count255}
 \def\Roman#1{\uppercase\expandafter{\romannumeral #1}}
 \def\roman#1{\romannumeral #1}
 
%%%%
%%%%           PAGING CONTROL MACROS
%%%%    INCLUDING HEADLINES AND FOOTLINES
%%%%
 
 \countdef\pagenumber=1  \pagenumber=1
 \def\advancepageno{\global\advance\pageno by 1
    \ifnum\pagenumber<0 \global\advance\pagenumber by -1
     \else\global\advance\pagenumber by 1 \fi \global\frontpagefalse }
\def\pagefolio#1{\ifnum#1<0 \romannumeral-#1
            \else \number#1 \fi }
 
\def\folio{\pagefolio{\pagenumber}}
 \def\pagecontents{
    \ifvoid\topins\else\unvbox\topins\vskip\skip\topins\fi
    \dimen@ = \dp255 \unvbox255
    \ifvoid\footins\else\vskip\skip\footins\footrule\unvbox\footins\fi
    \ifr@ggedbottom \kern-\dimen@ \vfil \fi }
 \def\makeheadline{\vbox to 0pt{ \hfuzz=30pt \skip@=\topskip
       \advance\skip@ by -12pt \advance\skip@ by -2\normalbaselineskip
       \vskip\skip@ \line{\vbox to 12pt{}\the\headline\hfill} \vss
       }\nointerlineskip}
 \def\makefootline{\baselineskip = 1.5\normalbaselineskip
                  \line{\the\footline}}
 \def\nopagenumbers{\p@genumfalse}
 \def\pagenumbers{\p@genumtrue}
 
%%% Listing break command expansion
 
 \def\tableconbreak{\break}
 \def\tableconbreakspace{\ }
 \def\tableconbreakfill{\hfill\break}

%% listing break command
 
 \let\conbreak=\tableconbreak
 \let\tbreak=\tableconbreakspace
 
%%%%%
%%%%% CHAPTER, SECTION, SUBSECTION, ITEM AND POINT MACROS
%%%%%
 
%%%
%%%         SET UP PARAMETERS FOR PAPER DIVISIONS
%%%
 
 \def\splitprep{\global\newlinechar=`\^^J}
 \def\splitprepend{\global\newlinechar=-1}

 \def\consection#1{
    \iftabl@conlist
       \splitprep
       \immediate\write\tableconwrite{\string\immediate
                                      \string\spacecheck\sectionminspace
                                     }
       \immediate\write\tableconwrite{\string\vskip0.25in
                         \string\titlestyle{\string\bf\ #1 }%
                         \string\vskip0.0in
                         \string\nullbox{1pt}{1pt}{1pt}%
                                     }
       \splitprepend
    \fi}
 \def\unnumberedchapters{\let\makel@bel=\relax \let\chapterlabel=\relax
             \let\sectionlabel=\relax \let\subseclabel=\relax \equanumber=-1 }
 \def\titlestyle#1{\par\begingroup \interlinepenalty=9999
      \leftskip=0.02\hsize plus 0.23\hsize minus 0.02\hsize
      \rightskip=\leftskip \parfillskip=0pt
      \hyphenpenalty=9000 \exhyphenpenalty=9000
      \tolerance=9999 \pretolerance=9000
      \spaceskip=0.333em \xspaceskip=0.5em
      \iftwelv@\fourteenpoint\else\twelvepoint\fi
    \noindent #1\par\endgroup }
 \def\spacecheck#1{\p@gecheck{#1}%
    \ifsp@cecheck
       \else \vfill\eject
    \fi}
 \def\majorreset{
    \ifnum\figurecount<0\else\global\figurecount=1\fi
    \ifnum\equanumber<0 \else\global\equanumber=1\fi
    \sectionnumber=0 \subsectionnumber=0
    \tablecount=0  \problemnumber=0
    \bookheadline={}%
    \chapterheadline={}%
    }
 \def\chapterreset{\global\advance\chapternumber by 1
                   \majorreset
                   \makel@bel}
 \def\appendflag#1{
        \if@ppendix
        \else
          \global\@ppendixtrue
          \starttable{APPENDICES CALLED}{\appendixout}{\appendixwrite}{6}
        \fi
        \@ddconentry{\appendixwrite}{\noindent{\bf #1}}{1}
                   }
 
%%%
%%%               CHAPTER MACRO
%%%
 
 \def\chapter#1{\penalty-300%
    \spacecheck\chapterminspace%
    \chapterreset%
    \ifIEEE%
       \chapterheadline={\tenbf\noindent{\chapterlabel}.\quad #1}%
       \else%
       \chapterheadline={\tenrm\chapterlabel\quad #1}%
      \fi%
    \bookheadline={\tenrm \chapterlabel\quad #1}%
    \ifb@@kstyle%
         \global\FIRSTP@GE%
         \ifodd\pagenumber%
            \else%
               \par\nullbox{1pt}{1pt}{1pt}%
               \vfill\supereject%
         \fi%
         \global\FIRSTP@GE%
         \vtop{\line{\hfill{\ninebf %
               CHAPTER}\hskip12pt{\twentyfourbf\the\chapternumber}}%
         \vglue0.5in%
         \baselineskip=26pt%
         \let\tbreak=\par%
         \let\conbreak=\par%
           \bgroup \everypar={\hfill}%
                   {\twentyfourpoint \bf#1}%
           \egroup}%
         \let\conbreak=\tableconbreakspace%
         \let\tbreak=\tableconbreakspace%
         \bookstyle%
         \vglue\bookchapterskip%
       \else%
          \ifIEEE%
            \par\nobreak\vskip\headskip%
            \vskip\chapterskip%
            \noindent\bf{{\chapterlabel}\  #1}\rm
          \else%
            \par\n@ll%
            \nobreak\vskip\headskip%
            \vskip\chapterskip%
            \titlestyle{\chapterlabel \ #1}%
          \fi%
    \fi%
    \nobreak\vskip\headskip%
    \let\conbreak=\tableconbreak%
    \iftabl@conlist%
      \@ddconentry{\tableconwrite}{\hskip0.0pt plus 0pt minus 0pt
                                   \chapterlabel\quad\ #1}{1}%
    \fi%
    \penalty 30000%
    \wlog{\string\chapter\ \chapterlabel} }%
 %
 
%%%
%%%            SECTION MACRO
%%%
 
 \def\section#1{\par \let\conbreak=\tableconbreakfill%
    \ifnum\the\lastpenalty=30000\else%
    \penalty-200\vskip\sectionskip \immediate\spacecheck\sectionminspace\fi%
    \subsectionnumber=0 %reset the subsection numbering%
    \wlog{\string\section\ \chapterlabel \the\sectionnumber}%
    \global\advance\sectionnumber by 1  \noindent%
    \bookheadline={\tenrm \chapterlabel\sectionlabel\quad #1}%
    \ifb@@kstyle
        {\baselineskip=9.5pt%
        \setbox0=\hbox{\twelvebf\chapterlabel
                         \the\sectionnumber\ }%
        \let\conbreak=\par%
        \let\tbreak=\par%
        \bgroup \everypar={\ }%
           \hskip-\wd0{\twelvepoint\bf\box0\hskip\parindent#1}
        \egroup\par}%
        \let\tbreak=\tableconbreakspace%
        \let\conbreak=\tableconbreakfill%
      \else%
         \ifIEEE%
           \setbox0=\hbox{\it{\chapterlabel\sectionlabel}.}%
           \noindent{\it\box0\quad #1}\par\rm%
         \else%     
           \setbox0=\hbox{\caps\chapterlabel \sectionlabel\ }%
           \noindent{\caps\box0\quad #1}\par%
         \fi%
    \fi
    \nobreak\vskip\headskip%
    \let\conbreak=\tableconbreakfill%
    \iftabl@conlist%%
      \@ddconentry{\tableconwrite}{%
            \hskip0.5in\ \chapterlabel\sectionlabel \quad\ #1}{2}%
    \fi%
    \penalty 30000 }%
 
%%%
%%%         SUBSECTION MACRO
%%%
 \def\subsection#1{\let\conbreak=\tableconbreakfill%
    \par%
    \ifIEEE\subsecstyle={\number}\fi%
    \global\advance\subsectionnumber by 1%
    \ifnum\the\lastpenalty=30000\else \penalty-100\smallskip \fi%
    \wlog{\string\subsection\ \chapterlabel\the\sectionnumber \ %
           \subseclabel}%
    \iftabl@conlist%
       \@ddconentry{\tableconwrite}{%
              \hskip1.0in\ \chapterlabel\sectionlabel\ %
              \subseclabel\quad #1}{3}%
    \fi%
    \ifIEEE%
       \noindent\it\chapterlabel\sectionlabel{.}\the\subsectionnumber{.}\quad%
       \it{#1}\enspace \vadjust{\penalty5000}\rm\par%
     \else
       \noindent\chapterlabel\sectionlabel \ \subseclabel\quad%
       \undertext{#1}\enspace \vadjust{\penalty5000}\fi}%
 %
 
%%%
%%%            APPENDIX MACROS
%%%
 
 \def\APPENDIX#1#2{%
    \par\penalty-300\vskip\chapterskip%
    \spacecheck\chapterminspace \majorreset%
    \nobreak\vskip\headskip%
    \ifb@@kstyle
         \global\FIRSTP@GE%
         \par\nullbox{1pt}{1pt}{1pt}
         \ifodd\pagenumber% do nothing
            \else \par\nullbox{1pt}{1pt}{1pt}
                  \vfill\supereject            % get rid of even page
         \fi
         \global\FIRSTP@GE%
         \vtop{\line{\hfill{\ninebf %
               APPENDIX}\hskip12pt{\twentyfourbf #1}}%
         \vglue0.5in%
         \baselineskip=26pt%
         \let\tbreak=\par%
         \let\conbreak=\par%
           \bgroup \everypar={\hfill}%
                   {\twentyfourpoint \bf#2}%
           \egroup}%
         \let\conbreak=\tableconbreakspace%
         \let\tbreak=\tableconbreakspace%
         \bookstyle%
         \vglue\bookchapterskip%
      \else
          \immediate\titlestyle{\bf APPENDIX {\bf #1} \break%
                                   {\bf #2}}%
    \fi
    \nobreak\vskip\headskip \penalty 30000%
    \xdef\chapterlabel{\appendixlabel .}%
    \bookheadline={\tenrm \chapterlabel\quad #2}%
    \chapterheadline={\tenrm \chapterlabel\quad #2}%
    \iftabl@conlist%
       \iffirst@ppendix%
           \immediate\write\tableconwrite{%
                        \string\immediate
                        \string\spacecheck\sectionminspace
                                         }%
           \immediate\write\tableconwrite{\nullbox{1pt}{1pt}{1pt}%
                                         }%
           \immediate\write\tableconwrite{\vskip 0.375in
                                          \centerline{\fourteenrm\bf
                                                       APPENDICES}%
                                         }%
           \immediate\write\tableconwrite{\vskip\headskip
                                         }%
           \global\first@ppendixfalse%
       \fi%
       \@ddconentry{\tableconwrite}{%
             {}\chapterlabel\quad\  #2}{1}%
    \fi%
    \wlog{\string\Appendix\ \chapterlabel} }%
 \def\Appendix#1{\APPENDIX{#1}{}}%
 \def\appendix#1{%
      \global\advance\appendixnumber by 1%
      \xdef\appendixlabel{\the\appendixstyle{\the\appendixnumber} }%
      \APPENDIX{\appendixlabel}{#1}}%
 
%%%
%%%           PROBLEM MACROS
%%%
 
 \def\Problems{\par\n@ll \penalty-300 \vskip\chapterskip
    \global\problemnumber=0
    \spacecheck\chapterminspace
    \bookheadline={\tenrm  Problems}%
    \nobreak\vskip\headskip%
    \noindent\hskip-0.475in{\twelvepoint\bf\ Problems}%
    \nobreak\vskip\headskip%
    \nobreak\vskip\headskip%
    \iftabl@conlist
       \@ddconentry{\tableconwrite}{%
             \hskip0.563in Problems}{2}%
    \fi
    \penalty 30000
    \wlog{CHAPTER PROBLEMS}%
    }
 
 \def\problem{\global\advance\problemnumber by 1
         \chapterlabel\the\problemnumber :\quad}
 
%%%
%%%          FURTHER STUDY MACRO
%%%
 
 \def\Furtherstudy{\par\n@ll \penalty-300 \vskip\chapterskip%
    \spacecheck\chapterminspace%
    \bookheadline={\tenrm Suggestions for Further Study}%
    \noindent\hskip-0.475in{\twelvepoint\bf\ Suggestions for Further Study}%
    \nobreak\vskip\headskip%
    \iftabl@conlist%
      \@ddconentry{\tableconwrite}{%
             \hskip0.563in Suggestions for Further Study}{2}%
    \fi%
    \nobreak\vskip\headskip%
    \penalty 30000%
    \wlog{FURTHER STUDY} }%
 
%%%
%%%            INDEX MACROS
%%%

\def\indexoff{\indexfalse
              \closeout\indexwrite}
\def\indexing#1{%% Index call allowing a comment
     \ifindex%
        \conpage=\pagenumber
        \p@gecheck{0pt}%
        \ifsp@cecheck
           \else \advance\conpage by 1 %advance relative page by one
        \fi
        \immediate\write\indexwrite{#1 \the\conpage
                                   }%
     \fi}%

\indexoff %% default setting
 
%%%
%%%             ACKNOWLEDGEMENTS MACRO
%%%

%%%
%%%           ITEM AND POINT MACRO
%%%
 
 \def\Textindent#1{\noindent\llap{#1\enspace}\ignorespaces}
 \def\GENITEM#1;#2{\par \hangafter=0 \hangindent=#1
     \Textindent{$ #2 $}\ignorespaces}
 \outer\def\newitem#1=#2;{\gdef#1{\GENITEM #2;}}
 \newdimen\itemsize                \itemsize=30pt
 \newitem\item=1\itemsize;
 \newitem\sitem=1.75\itemsize;     
 \newitem\ssitem=2.5\itemsize;     
 
 \outer\def\newlist#1=#2&#3&#4;{\toks0={#2}\toks1={#3}%
    \count255=\escapechar \escapechar=-1
    \alloc@0\list\countdef\insc@unt\listcount     \listcount=0
    \edef#1{\par
       \countdef\listcount=\the\allocationnumber
       \advance\listcount by 1
       \hangafter=0 \hangindent=#4
       \Textindent{\the\toks0{\listcount}\the\toks1}}
    \expandafter\expandafter\expandafter
     \edef\c@t#1{begin}{\par
       \countdef\listcount=\the\allocationnumber \listcount=1
       \hangafter=0 \hangindent=#4
       \Textindent{\the\toks0{\listcount}\the\toks1}}
    \expandafter\expandafter\expandafter
     \edef\c@t#1{con}{\par \hangafter=0 \hangindent=#4 \noindent}
    \escapechar=\count255}
 \def\c@t#1#2{\csname\string#1#2\endcsname}
 \newlist\point=\Number&.&1.0\itemsize;
 \newlist\subpoint=(\alphabetic&)&1.75\itemsize;
 \newlist\subsubpoint=(\roman&)&2.5\itemsize;

 \newskip\separationskip   \separationskip=\parskip
 
%%%%%%%
%%%%%%%              LISTING MACROS
%%%%%%%
 
%%%  Paramemeters
 
 \newtoks\temphold
 \newtoks\ta
 \newtoks\tb
 \newtoks\captiontoks
 \newwrite\capwrite
 %\immediate\openout\capwrite=cap_test.tex %%%flag
 \newcount\runner  \runner=0
 \newcount\wordsnum \wordsnum=0
\newif\iflongc@p  \longc@pfalse
\def\longcap{\bgroup\obeyspaces\endlinechar=-1%\catcode`\^^M=\active %
             \global\longc@ptrue}
\def\endlongcap{\egroup\global\longc@pfalse}
 
 \def\confolio{\pagefolio{\conpage}}
 \def\p@gecheck#1{
        \dimen@=\pagegoal
        \advance\dimen@ by -\pagetotal
        \ifdim\dimen@<#1
           \global\sp@cecheckfalse
           \else\global\sp@cechecktrue
        \fi}
 \def\st@rtt@ble#1#2#3{
                 \message{listing #1:  type
                          \noexpand#2 for list}
                 \global\setbox#3=\vbox{\normalbaselines
                      \titlestyle{\seventeenrm\bf #1} \vskip\headskip}}
 \def\starttable#1#2#3#4{%
     \message{external listing of #1:  type %
              \noexpand#2 for list}%
     \ifcase#4%
           \immediate\openout#3=TABLE_OF_CONTENTS.TEX%  % 0 first list
       \or%
           \immediate\openout#3=FIG_CAP_AND_PAGE.TEX%   % 1 USER CAPTIONS
       \or%
           \immediate\openout#3=FIGURE_CAPTIONS.TEX%    % 2 PUBLICATION CAPTIONS
       \or%
           \immediate\openout#3=TABLE_CAPTIONS.TEX%     % 3 PUBLICATION CAPTIONS
       \or%
           \immediate\openout#3=TAB_CAP_AND_PAGE.TEX%   % 4 USER CAPTIONS
       \or%
           \immediate\openout#3=EQN_PAGE.TEX%           % 5 USER EQUATIONS
       \or%
           \immediate\openout#3=APP_CALLED.TEX%         % 6 APPENDICES CALLED
       \or%
           \immediate\openout#3=REFERENC.TEX%           % 7 REFERENCES
      \else \immediate\message{You are in big trouble call a %
                                 TeXnician}%
     \fi%
     \ifIEEE
        \ifnum#4=7\immediate\write#3{\string\vbox {\string\normalbaselines%
              \string\bf\noindent References \string\vskip \string\headskip}} 
        \fi%
     \else\immediate\write#3{\string\vbox {\string\normalbaselines%
               \string\titlestyle {\string\seventeenrm \string\bf \ #1}
               \string\vskip \string\headskip}}\fi%
 }%

\def\untoken#1#2{% #1 is the tok name, #2 are the new tokens to add
     \ta=\expandafter{#1}\tb=\expandafter{#2}%
     \immediate\edef\temphold{\the\ta\the\tb}
     \global\captiontoks=\expandafter{\temphold}}%
 
\def\getc@pwrite#1{
                   \ifx#1\endcaption%
                      \let\next=\relax%
                      \immediate\write\capwrite{\the\captiontoks}
                      \global\captiontoks={}
                    \else%
                       \untoken{\the\captiontoks}{#1}
                       \ifx#1\space%
                             \advance\runner by 1
                             \advance\wordsnum by1 \let\next=\getc@pwrite
                       \fi%
                    \let\next=\getc@pwrite\fi%
                 \ifnum\runner=10
                       \immediate\message{.}
                       \immediate\write\capwrite{\the\captiontoks}
                       \global\captiontoks={}
                       \runner=0
                 \fi
                       \next}%
 
\def\c@pwrite#1#2{
               \let\capwrite=#1
               \global\runner=0
               \immediate\message{working on long caption. .}
               \wordsnum=0 \getc@pwrite#2\endcaption
               \immediate\message{The number of words= \number\wordsnum}}

\def\t@@bl@build@r#1#2{\relax%              #1 multiplier number
         \let\conbreak=\tableconbreakfill % #2 page number
         \conpage=#2
         \t@mp=\tablelineskip
         \multiply\t@mp by #1%%
         \@@@@@=\tablelinelength \advance\@@@@@ by -\t@mp%
                                 \advance\@@@@@ by -1in%
             \@@@@@two=\tablelinelength%
             \advance\@@@@@two by -1in%
             \vglue\separationskip
         \vbox\bgroup
             \vbox\bgroup\parshape=3 0pt\@@@@@two%
                               \t@mp\@@@@@%
                               \t@mp\@@@@@%
                   \parfillskip=0pt%
                   \pretolerance=9000 \tolerance=9999 \hbadness=2000%
                   \hyphenpenalty=9000 \exhyphenpenalty=9000%
                   \interlinepenalty=9999%
                   \tablespace
                   \vskip\baselineskip\raggedright\noindent}
%Place the caption between these commands
\def\endt@@bl@build@r{{\tabledotfill \hskip-0.07in $\,$}%
                  \egroup%
             \parshape=0%
            \vskip\tabledotvskip\hskip\@@@@@two\hskip\tabledothskip%
            { \tabledotfill\quad} {\confolio}%
         \egroup
         \vskip-0.10in}%
 
 \def\addconentry#1#2{%
     \setbox0=\vbox{\normalbaselines #2}%
     \relax%
     \global\setbox#1=\vbox{\unvbox #1 \vskip 4pt plus 2pt minus 1pt \box0 }}%

\def\@ddconentry#1#2#3{%  #1 where to write
                       %  #2 what
                       %  #3 number
     \conpage=\pagenumber%
     \p@gecheck{0pt}%
     \ifsp@cecheck \else
         \if\conpage<0
               \advance\conpage by -1
            \else
               \advance\conpage by 1
         \fi%
     \fi%
     \let\conbreak=\tableconbreakfill
     \splitprep
     \immediate\write#1
              {\string\t@@bl@build@r}%
     \immediate\write#1{{#3}%
                        {\the\conpage}%
                       }%
     \iflongc@p
         \immediate\c@pwrite{#1}{ #2 }%
       \else
         \immediate\write#1{{#2}}%
     \fi
     \immediate\write#1
              {\string\endt@@bl@build@r}%
     \splitprepend
}

\def\@ddfigure#1#2#3#4{% #1 -> where to write
                       % #2 -> figure/table number
                       % #3 -> caption
                       % #4 -> Figure/Table Tag
        \splitprep
        \captiontoks={}
        \immediate\write#1{{\string\vglue\string\separationskip}%
                          }%
        \immediate\write#1
              {\string\par \string\nullbox {1pt}{1pt}{1pt}%
              }%
        \immediate\write#1{\string\vbox}%
        \immediate\write#1{\string\bgroup \string\tablespace}
        \immediate\write#1{\string\parshape=2}
        \immediate\write#1{0pt\string\tablelinelength}
        \immediate\write#1{\string\figurelineskip\string\@@@@@}
        \immediate\write#1{\string\bgroup
                           }
       \iflongc@p
            \immediate\c@pwrite{#1}{#4~#2:\quad\ #3}
          \else
            \immediate\write#1{#4~#2:\quad\ #3}
       \fi
       \immediate\write#1{\string\egroup
                          \string\egroup
                         }%
       \immediate\write#1{\string\parshape=0%
                          }%
       \splitprepend
                     }%
 
 \def\captionsetup{
        \@@@@@=\tablelinelength \advance\@@@@@ by -\figurelineskip}
 \def\manualpageno{\manualpagenotrue}
 
 \def\dumplist#1{%
     \bookheadline={}%
     \chapterheadline={}%
     \unlock   % allow command characters
     \global\let\conbreak=\tableconbreakfill
     \ifmanualpageno
         \else
            \pagenumber=1
     \fi
     \ifb@@kstyle
        \global\multiply\pagenumber by -1
        \bookheadline={}
     \fi
           %% count in roman numerals if in book format
           %% count in arabic numerals if in paper format
     \vskip\chapterskip  %% skip down page the distance of a chapter title
     \ifcase#1
            \chapterheadline={Contents}
            \bookheadline={Contents}    %      \conout
            \input TABLE_OF_CONTENTS.TEX%      0 list generated this run
       \or
            \chapterheadline={Contents}
            \bookheadline={Contents}         % \Conout
            \input User_table_of_contents.tex% 1 list generated from other runs
       \or
            \chapterheadline={Contents}
            \bookheadline={Contents}         % \CONOUT
            \input User_table_of_contents.tex% 2 list generated from other runs
            \input TABLE_OF_CONTENTS.TEX%        list generated from this run
       \or
           {\obeylines
            \input index.tex%                   3
           }
       \or
           {\obeylines
            \input User_index.tex%              4
           }
       \or
           {\obeylines
            \input User_index.tex%              5
            \input index.tex%
           }
       \or
           \captionsetup
           \input FIGURE_CAPTIONS.TEX%          6
       \or
           \input FIG_CAP_AND_PAGE.TEX%         7
       \or
           \captionsetup
           \input TABLE_CAPTIONS.TEX%           8
       \or
           \input TAB_CAP_AND_PAGE.TEX%         9
       \or
           \input EQN_PAGE.TEX%                10
       \or
           \input APP_CALLED.TEX%              11
       \or
           \input REFERENC.TEX%                12
       \else \immediate\message{ Call a TeXnician, You are in BIG trouble}
     \fi
     \vfill\eject %% fill the page and start a new one
     \lock}
 \def\dumpbox#1{         %% assume that last page is filled
     \ifb@@kstyle
        \multiply\pagenumber by -1
        \bookheadline={}
     \fi
           %% count in roman numerals if in book format
           %% count in arabic numerals if in paper format
     \vskip\chapterskip  %% skip down page the distance of a chapter title
     \unvbox#1           %% download the list
     \vfill\eject}       %% fill the page and start a new one
 
%%%%
%%%%     LIST GENERATORS
%%%%
 
 \def\notabledots{\global\let\tabledotfill=\hfill}
 \def\tabledots{\global\let\tabledotfill=\dotfill}
 \def\tableconlist{\global\tabl@conlisttrue
                   \starttable{CONTENTS}{\conout}{\tableconwrite}
                              {0}}
 \def\tableconlistoff{\global\tabl@conlistfalse
                      \immediate\closeout\tableconwrite}
 
 %%% REFERENCE LIST GENERATED IN REFERENCE ERROR CHECKING
 \def\reflist{\global\referencelisttrue}

 \def\figurelist{\global\figur@listtrue
                 \starttable{FIGURE CAPTIONS AND PAGES}
                       {\figuresout}{\figurewrite}{1}
                 \starttable{FIGURE CAPTIONS}{\figout}{\figwrite}{2}
                }
 \def\figurelistoff{\global\figur@listfalse}

 \def\tablelist{\global\tabl@listtrue
                \starttable{TABLE CAPTIONS}{\tabout}{\tablewrite}{3}
                \starttable{TABLE CAPTIONS AND PAGES}
                           {\tablesout}{\tableswrite}{4}
               }
 \def\tablelistoff{\global\tabl@listfalse}
 
%%%
%%%       EQUATIONS WITH PAGE NUMBERS
%%%
 
 \def\equationlist{\global\@qlisttrue
                 \starttable{EQUATIONS}{\eqout}{\eqnwrite}{5}
                 \immediate\message{Producing an external list of equation
                         names in INTERNAL_EQUATION_LIST.TEX}
                 \immediate\openout\equ@tions=internal_equation_list.tex
                  }
 
 \def\equationlistoff{\global\@qlistfalse}

%%%
%%%       MAKE LOADABLE LIST OF NAMES
%%%
 
 \def\produceequations#1{
     \ifnum\equanumber<0
          \immediate\write\equ@tions{\string\xdef \string#1
                  {{\string\rm  (\number-\equanumber)}} }
       \else
          \immediate\write\equ@tions{\string\xdef \string#1
                  {{\string\rm  (\chapterlabel\number\equanumber)}} }
     \fi}
 
%%%
%%%    LOADING MACRO FOR LIST
%%%
 
 \def\equ@tionlo@d{ \ifeqlo@d
                       \else \input BOOK_EQUATIONS
                             \eqlo@dtrue
                       \fi}
 
%%%
%%%           DOWNLOADING THE LISTS
%%%

 \def\appendixout{\immediate\closeout\appendixwrite
                  \dumplist{11}
                 }
 \def\conout{\normalspace\immediate\closeout\tableconwrite
             \dumpbox{\tableconbox}
             \dumplist{0}}

 \def\refout{\immediate\closeout\referencewrite
             \dumplist{12}
%   \par \penalty-400 \vskip\chapterskip
%   \spacecheck\referenceminspace \immediate\closeout\referencewrite
%   \referenceopenfalse
%   \line{\ifIEEE\bf\hfil References\hfil
%         \else\fourteenrm\hfil REFERENCES\hfil\fi%}
%     \vskip\headskip
%   \input referenc.tex
   }
 \def\figout{%\dumpbox{\figbox}
             \immediate\closeout\figwrite
             \dumplist{6}}
 \def\figuresout{%\dumpbox{\figurebox}
             \normalspace\immediate\closeout\figurewrite
             \dumplist{7}}
 
 \def\tabout{\immediate\closeout\tablewrite
             \dumplist{8}}
 \def\tablesout{\normalspace\immediate\closeout\tableswrite
          \dumplist{9}}
 
 \def\eqout{\immediate\closeout\eqnwrite
          \dumplist{10}
          \immediate\closeout\equ@tions}
 
%%%%%%
%%%%%%           REFERENCE MACROS
%%%%%%
 
%%%
%%%           SET UP PARAMETERS
%%%
 
\newdimen\referenceminspace  \referenceminspace=25pc
\newcount\referencecount     \referencecount=0
\newcount\titlereferencecount     \titlereferencecount=0
\newcount\lastrefsbegincount \lastrefsbegincount=0
\newdimen\refindent     \refindent=30pt
\newif\ifreferencelist       \global\referencelisttrue
\newif\ifreferenceopen       \global\referenceopenfalse
\newwrite\referencewrite
\newwrite\equ@tions
\newtoks\rw@toks
 
%%%
%%%           REFERENCE MARKING MACROS
%%%
 
\def\NPrefmark#1{\attach{\scriptscriptstyle [ #1 ] }}
\def\PRrefmark#1{\attach#1}
\def\IEEErefmark#1{ [#1]}
\def\refmark#1{\relax%
     \ifPhysRev\PRrefmark{#1}%
     \else%
        \ifIEEE\IEEErefmark{#1}%
          \else%
            \NPrefmark{#1}%
          \fi%
     \fi}
 
%%%%%
%%%%%            THE REFERENCE MACROS
%%%%%

\def\REF#1#2{\space@ver{}\refch@ck
   \global\advance\referencecount by 1 \xdef#1{\the\referencecount}%
   \r@fwrit@{#1}{#2}}%
\def\ref#1{\REF\?{#1}\refend}%
\def\refend{\global\lastrefsbegincount=\referencecount
            \refsend}%                                 SINGLE REF
\def\refsend{\refmark{\count255=\referencecount %    %  MULTIPLE REFS
   \advance\count255 by-\lastrefsbegincount %
   \ifcase\count255 %
       \number\referencecount%                      % single reference
   \or%
      \number\lastrefsbegincount,\number\referencecount% two refs
   \else%
      \number\lastrefsbegincount -\number\referencecount% more than two refs
  \fi}}%
 
%%%
%%%        ERROR CHECKING
%%%
 
\def\refch@ck{\ifreferencelist%
                  \ifreferenceopen%
                       \else \global\referenceopentrue%
                           \starttable{REFERENCES}{\refout}{\referencewrite}{7}
                  \fi%
              \fi%
}
 
%%%
%%%        REFERENCE WRITING
%%%
 
\def\rw@begin#1\splitout{\rw@toks={#1}\relax%
   \immediate\write\referencewrite{\the\rw@toks}\futurelet\n@xt\rw@next}%
\def\rw@next{\ifx\n@xt\rw@end \let\n@xt=\relax%
      \else \let\n@xt=\rw@begin \fi \n@xt}%
\let\rw@end=\relax%
\let\splitout=\relax%
\def\r@fwrit@#1#2{%
   \splitprep%
   \immediate\write\referencewrite{\ifIEEE\noexpand\refitem{\noindent[#1]}
                                   \else\noexpand\refitem{#1.}\fi}%
   \rw@begin #2\splitout\rw@end \@sf%
   \splitprepend%
                 }%
 
%%%
%%%          REFERENCE LABELING
 
\def\refitem#1{\par \hangafter=0 \hangindent=\refindent \Textindent{#1}}%
 
%%%%%
%%%%%          TITLE REFERENCES
%%%%%
 
%%%
%%%            TITLE REFERENCE MARKING
%%%
 
%
%
 
%%%%
%%%%          THE MACRO
%%%%
 
\def\TITLEREF#1#2{\space@ver{}\refch@ck%
   \global\advance\titlereferencecount by 1%
   \xdef#1{\alphabetic{\the\titlereferencecount}}%
   \r@fwrit@{#1}{#2}}%
%
 
%%%%%%%
%%%%%%%       FIGURE MACROS
%%%%%%%
 
%%%
%%%       ERROR CHEDKING
%%%
 
 \def\checkreferror{\ifinner\errmessage{This is a wrong place to DEFINE
     a reference or a figure! You may have your references / figure
     captions screwed up. }\fi}

%%%%%
%%%%%            THE MACROS
%%%%%
 
% \def\FIG#1#2{\checkreferror
%     \ifnum\figurecount<0
%          \xdef#1{\number-\figurecount}%
%          \advance\figurecount by -1
%       \else
%          \xdef#1{\chapterlabel\number\figurecount}%
%          \advance\figurecount by 1
%     \fi
%     \iffigur@list
%        \@@@@@=\hsize \advance\@@@@@ by -1.219in
%        \@ddconentry{\figurewrite}%
%            {\noindent{\bf#1.}\quad{\string\string\string#1}\quad\ #2}{1}%
%        \@ddfigure{\figwrite}{#1}{#2}{Figure}%
%     \fi     }%
 
 \def\FIG#1#2{\checkreferror
     \ifnum\figurecount<0
          \global\xdef#1{\number-\figurecount}%
          \global\advance\figurecount by -1
       \else
          \global\xdef#1{\chapterlabel\number\figurecount}%
          \global\advance\figurecount by 1
     \fi
     \iffigur@list
        \@@@@@=\hsize \advance\@@@@@ by -1.219in
        \@ddconentry{\figurewrite}%
            {\noindent
                \bgroup \bf \expandafter{\csname\string#1.\endcsname} \egroup
                \quad{#1}\quad\ #2
            }
            {1}%
        \@ddfigure{\figwrite}{#1}{#2}{Figure}%
     \fi     }%

%%%%%%%
%%%%%%%        TABLE MACROS
%%%%%%%
 
%%%
%%%       ACCOUNTING
%%%
 
 \def\nexttable{\checkreferror\global\advance\tablecount by 1}
 
%%%%%
%%%%%      THE MACROS
%%%%%
 
% \def\TABLE#1#2{\nexttable\xdef#1{\chapterlabel\the\tablecount}%
%     \iftabl@list
%        \@ddfigure{\tablewrite}{#1}{#2}{Table}%
%        \@ddconentry{\tableswrite}%
%               {\noindent{\bf#1.}\quad{\string\string\string#1}\quad\ #2}{1}%
%     \fi}%
 
 \def\TABLE#1#2{\nexttable\xdef#1{\chapterlabel\the\tablecount}%
     \iftabl@list
        \@ddfigure{\tablewrite}{#1}{#2}{Table}%
        \@ddconentry{\tableswrite}%
               {\noindent
                   \bgroup \bf \expandafter{\csname\string#1.\endcsname} \egroup
                   \quad{#1}\quad\ #2}
               {1}%
     \fi}%

%%%%%%%
%%%%%%%                 EQUATION MACROS
%%%%%%%
 
%%%%
%%%%       NAMING MACRO
%%%%
 
% \def\eqname#1{\relax
%     \if@qlist
%       \produceequations{#1}
%     \fi
%     \ifnum\equanumber<0
%           \global\xdef#1{{\rm(\number-\equanumber)}}
%           \global\advance\equanumber by -1
%     \else
%           \global\xdef#1{{\rm(\chapterlabel \number\equanumber)}}
%           \global\advance\equanumber by 1
%     \fi
%     \if@qlist
%       \@ddconentry{\eqnwrite}
%           {\noindent{\bf#1.}\quad{\string\string\string#1}\quad}{1}
%     \fi      }
 
 \def\eqname#1{\relax
     \if@qlist
       \produceequations{#1}
     \fi
     \ifnum\equanumber<0
           \global\xdef#1{{\rm(\number-\equanumber)}}
           \global\advance\equanumber by -1
     \else
           \global\xdef#1{{\rm(\chapterlabel \number\equanumber)}}
           \global\advance\equanumber by 1
     \fi
     \if@qlist
       \@ddconentry{\eqnwrite}
           {\noindent
             \bgroup \bf \expandafter{\csname\string#1\endcsname.} \egroup
             \quad{#1}\quad}
           {1}
     \fi      }
 
%%%
%%%          NORMAL MATH MODE
%%%

%%%
%%%          BOXED MATH MODE
%%%
 
 \def\eqinsert#1{\noalign{\dimen@=\prevdepth \nointerlineskip
    \setbox0=\hbox to\displaywidth{\hfil #1}
    \vbox to 0pt{\vss\hbox{$\!\box0\!$}\kern-0.5\baselineskip}
    \prevdepth=\dimen@}}

%%%
%%%        ALIGNED MATH MODE
%%%

%%%
%%%     EQUATION NUMBERING SCHEMES
%%%

 \def\globaleqnumbers{\relax\if\equanumber<0\else\global\equanumber=-1\fi}
 
%%%
%%%     FIGURE NUMBERING SCHEMES
%%%

 \def\globalfigurenumbers{\relax\if\figurecount<0\else\global\figurecount=-1\fi}

%%%%%%%
%%%%%%%       CONTROLLING MACROS FOR THE INSERTION OF UG (and other postscript)
%%%%%%%       figures.
%%%%%%%

 \def\UGbody#1#2#3#4{
%      \special {ps::[asis, begin]      These are the old, ArborText specials
%         0 SPB
%         save
%           /showpage {} def
%           /initgraphics {} def
%           Xpos Ypos translate
%           #4 dup scale
%           #2 72 mul neg #3 72 mul neg translate
%      }
%      \special{ps: plotfile #1 asis}
%      \special{ps::[asis,end]
%          restore
%          0 SPE
%      }
    \UGdscale=#4\UGabs \multiply\UGdscale by 100 
    \UGscale=\number\UGdscale \divide\UGscale by \number\UGabs
    \UGdleft=#2\UGabs \multiply\UGdleft by 72
    \UGleft=\number\UGdleft \divide\UGleft by \number\UGabs 
            \advance\UGleft by 5            %Fudge it just a little....
    \UGdbot=#3\UGabs \multiply\UGdbot by 72 
    \UGbot=\number\UGdbot \divide\UGbot by \number\UGabs
%                                           %Scale the offsets....
    \multiply\UGbot by \number\UGscale  \divide\UGbot by 100
    \multiply\UGleft by \number\UGscale \divide\UGleft by 100
    \includegraphics{#1}
    \special {ps::[end] restore }
 }
 \def\UGfigs{
   %
% UGinsert parameters:
%	#1: name of UG generated postscript file
%	#2: figure width (inches)
%	#3: figure height (inches)
%	#4: left margin of figure (inches)
%	#5: bottom margin of figure (inches)
%	#6: desired figure magnification (1.0 = unmagnified)
%
% Parameters 2-5 refer to the dimensions when the UG generated
% PostScript file is printed as-is.
%
\def\UGinsert#1#2#3#4#5#6{
 \dimen0=#3in%
 \vbox to #6\dimen0{
 \vss
   \hbox to \hsize{%
     \hss
     \dimen0=#2in%
     \hbox to #6\dimen0{%
     \UGbody{#1}{#4}{#5}{#6}
     \hss
    }%
    \hss
   }
   \hss
  }
}
%
% The UGbody macro is included here for completeness, although it is now
% part of PHYZZM
%
% \def\UGbody#1#2#3#4{
%      \special {ps::[asis, begin]
%         0 SPB
%         save
%           /showpage {} def
%           /initgraphics {} def
%           Xpos Ypos translate
%           #4 dup scale
%           #2 72 mul neg #3 72 mul neg translate
%      }
%      \special{ps: plotfile #1 asis}
%      \special{ps::[asis,end]
%          restore
%          0 SPE
%      }
% }

   % Insert a UG figure with the figure caption on the right side of the
% figure.
%
% UGrcaption parameters:
%	#1: name of UG generated postscript file
%	#2: figure width (inches)
%	#3: figure height (inches)
%	#4: left margin of figure (inches)
%	#5: bottom margin of figure (inches)
%	#6: desired figure magnification (1.0 = unmagnified)
%	#7: figure caption
%
% Parameters 2-5 refer to the dimensions when the UG generated
% PostScript file is printed as-is.
%
\def\UGrcaption#1#2#3#4#5#6#7{
  \dimen0=#3in%
  \vbox to #6\dimen0{
    \dimen0=#2in
    \leftskip=#6\dimen0
    #7 \hfill
    \vss
    \UGbody{#1}{#4}{#5}{#6}
  }
}
%
% The UGbody macro is included here for completeness, although it is now
% part of PHYZZM
%
% \def\UGbody#1#2#3#4{
%      \special {ps::[asis, begin]
%         0 SPB
%         save
%           /showpage {} def
%           /initgraphics {} def
%           Xpos Ypos translate
%           #4 dup scale
%           #2 72 mul neg #3 72 mul neg translate
%      }
%      \special{ps: plotfile #1 asis}
%      \special{ps::[asis,end]
%          restore
%          0 SPE
%      }
% }

   % Version for dvi2ps driver
%
% Insert 2 UG drawings side by side
%
% UGtwoinsert parameters:
%	#1: name of left UG generated postscript file
%	#2: name of right UG generated postscript file
%	#3: figure width (inches)
%	#4: figure height (inches)
%	#5: left margin of figure (inches)
%	#6: bottom margin of figure (inches)
%	#7: desired figure magnification (1.0 = unmagnified)
%
% Parameters 3-6 refer to the dimensions when the UG generated
% PostScript file is printed as-is.  Both drawings must be the same
% size and have the same margins when printed as-is.
%
\def\UGtwoinsert#1#2#3#4#5#6#7{
  \dimen0=#4in%
  \vbox to #7\dimen0{
    \vss
    \hbox to \hsize{%
      \hss
      \dimen0=#3in%
      \hbox to #7\dimen0{%
      \UGbody{#1}{#5}{#6}{#7}
        \hss
      }%
      \hss
      \hbox to #7\dimen0{%
      \UGbody{#2}{#5}{#6}{#7}
        \hss
      }%
      \hss
    }
    \hss
  }
}
%
% The UGbody macro is included here for completeness, although it is now
% part of PHYZZM
%
% \def\UGbody#1#2#3#4{
%      \special {ps::[asis, begin]
%         0 SPB
%         save
%           /showpage {} def
%           /initgraphics {} def
%           Xpos Ypos translate
%           #4 dup scale
%           #2 72 mul neg #3 72 mul neg translate
%      }
%      \special{ps: plotfile #1 asis}
%      \special{ps::[asis,end]
%          restore
%          0 SPE
%      }
% }

   % Version for dvips driver
%
% Insert 2 UG drawings side by side
%
% UGtwocaption parameters:
%	#1: name of left UG generated postscript file
%	#2: name of right UG generated postscript file
%	#3: figure width (inches)
%	#4: figure height (inches)
%	#5: left margin of figure (inches)
%	#6: bottom margin of figure (inches)
%	#7: desired figure magnification (1.0 = unmagnified)
%	#8: left figure title
%	#9: right figure title
%
% Parameters 3-6 refer to the dimensions when the UG generated
% PostScript file is printed as-is.  Both drawings must be the same
% size and have the same margins when printed as-is.
%
\def\UGtwocapt#1#2#3#4#5#6#7#8#9{
  \dimen0=#4in%
  \vbox to #7\dimen0{
    \vss
    \hbox to \hsize{%
      \dimen0=#3in%
      \hbox to #7\dimen0{%
        \UGbody{#1}{#5}{#6}{#7}
        \hss
      }%
      \hss
      \hbox to #7\dimen0{%
        \UGbody{#2}{#5}{#6}{#7}
        \hss
      }%	
    }	
  }
  \smallskip
  \hbox to \hsize{%
    \dimen0=#3in%
    \setbox0=\hbox to #7\dimen0{\hsize=#7\dimen0\vtop{#8\hfill\vss}}%
    \setbox1=\hbox to #7\dimen0{\hsize=#7\dimen0\vtop{#9\hfill\vss}}%
    \box0 \hss \box1
  }
}
%
% The UGbody macro is included here for completeness, although it is now
% part of PHYZZM
%
% \def\UGbody#1#2#3#4{
%      \special {ps::[asis, begin]
%         0 SPB
%         save
%           /showpage {} def
%           /initgraphics {} def
%           Xpos Ypos translate
%           #4 dup scale
%           #2 72 mul neg #3 72 mul neg translate
%      }
%      \special{ps: plotfile #1 asis}
%      \special{ps::[asis,end]
%          restore
%          0 SPE
%      }
% }

 }

%%%%%%%
%%%%%%%       CONTROLLING MACRO DEFINITIONS FOR LETTERS
%%%%%%%
 
%%%
%%%                PARAMETERS
%%%
 
 \newskip\lettertopfil      \lettertopfil = 0pt plus 1.5in minus 0pt
 \newskip\spskip            \setbox0\hbox{\ } \spskip=-1\wd0
 \newskip\signatureskip     \signatureskip=40pt
 \newskip\letterbottomfil   \letterbottomfil = 0pt plus 2.3in minus 0pt
 \newskip\frontpageskip     \frontpageskip=1\medskipamount plus .5fil
 \newskip\headboxwidth      \headboxwidth= 0.0pt
 \newskip\letternameskip    \letternameskip=0pt
 \newskip\letterpush        \letterpush=0pt
 \newbox\headbox
 \newbox\headboxbox
 \newbox\physbox
 \newbox\letterb@x
 \newif\ifse@l              \se@ltrue
 \newif\iffrontpage
 \newif\ifletterstyle
 \newif\ifhe@dboxset        \he@dboxsetfalse
 \newtoks\memoheadline
 \newtoks\letterheadline
 \newtoks\letterfrontheadline
 \newtoks\lettermainheadline
 \newtoks\letterfootline
 \newdimen\holder
 \newdimen\headboxwidth
 \newtoks\myletterheadline  \myletterheadline={define \\myletterheadline}
 \newtoks\mylettername    \mylettername={{\twentyfourbf define}\ \\mylettername}
 \newtoks\phonenumber       \phonenumber={\rm (734) 764-4437}
 \newtoks\faxnumber	    \faxnumber={(734) 763-9694}
 \newtoks\telexnumber       \telexnumber={\tenfib 4320815 UOFM UI}
 
%%%%
%%%%          LETTER STYLE DEFINITIONS
%%%%
 
 \def\FIRSTP@GE{\ifvoid255\else\vfill\penalty-2000\fi
                \global\frontpagetrue}
 \def\FRONTPAGE{\ifvoid255\else\vfill\penalty-2000\fi
       \masterreset\global\frontpagetrue
       \global\lastp@g@no=-1 \global\footsymbolcount=0}
 
 \letterfootline={\hfil}
 \lettermainheadline={\hbox to \hsize{
        \rm\ifp@genum page \ \folio\fi
        \hfill\today}}
 \letterheadline{\hfuzz=60pt\iffrontpage\the\letterfrontheadline
      \else\the\lettermainheadline\fi}
 \def\addressee#1{\vskip-0.188in \singlespace
    \ialign to\hsize{\strut ##\hfill\tabskip 0pt plus \hsize \cr #1\hfill\crcr}
    \medskip\noindent\hskip\spskip}
 \def\letterstyle{\global\letterstyletrue%
                  \global\paperstylefalse%
                  \global\b@@kstylefalse%
                  \global\frontpagetrue%
                  \global\singlespace\global\lettersize}
 \def\lettersize{\hsize=6.5in\vsize=8.5in\hoffset=0in\voffset=0.250in
    \headboxwidth=\hsize  \advance\headboxwidth by -1.05in
    \skip\footins=\smallskipamount \multiply\skip\footins by 3}
 \def\telex#1{\telexnumber={\tenfib #1}}
 \def\noseal{\se@lfalse}
 
%%%
%%%         LETTER HEAD builders
%%%
 
 \def\myaddress#1{%
         \singlespace
         \global\he@dboxsettrue
         {\hsize=\headboxwidth \setbox0=\vbox{\ialign{##\hfill\cr #1 }}
         \global\setbox\headboxbox=\vbox {\hfuzz=50pt
                  \vfill
                  \line{\hfil\box0\hfil}}}}
 \def\MICH@t#1{\hfuzz=50pt\global\setbox\physbox=\hbox{\it#1}\hfuzz=1pt}
 \def\l@tt@rcheck{\ifhe@dboxset\else
                     \immediate\message{Setting default letterhead.}
                     \immediate\message{For more information consult the PHYZZM
                                       documents.}
                     \UM
                  \fi}
 \def\l@tt@rs#1{\l@tt@rcheck
              \vfill\supereject % Start a new page
              \global\letterstyle
              \global\letterfrontheadline={}
              \global\setbox\headbox=\vbox{{
                  \hbox to\headboxwidth{\hepbig\hfil #1\hfil}
                  \vskip 1pc
                  \unvcopy\headboxbox}}}
 \def\HEAD#1{\singlespace  %set spacing
    \global\letternameskip=22pt plus 0pt minus 0pt
    \global\memoheadline={UM\ #1\ Memorandum}
    \mylettername={\hep #1}
       \myaddress{
   \hep\hfill\hbox{The Harrison M. Randall Laboratory of Physics}\hfill \cr
   \hep\hfill\hbox{500 East University, Ann Arbor, Michigan 48109-1120}\hfill \cr
                                                 \cr}}
 
%%%%
%%%%            PUBLICATION TYPES
%%%%

% \def\PDK{\phonenumber={\rm (313) 764-4442}
%        \faxnumber={(313) 936-1817}
%        \HEAD{High Energy Physics}
%        \PUBNUM{PDK} \MICH@t{High Energy Physics}}
 \def\theory{\phonenumber={\rm (734) 763-9698}
        \HEAD{Theoretical Physics}
        \PUBNUM{TH} \MICH@t{Theoretical Physics}}
 \def\Eshop{\phonenumber={\rm (734) 763-1232}
        \faxnumber={(734) 936-6753}
        \HEAD{High Energy Physics Electronics Shop}
        \PUBNUM{HE} \MICH@t{High Energy Physics Electronics Shop}}
 \def\highenergy{\phonenumber={\rm (734) 764-4443}
        \faxnumber={(734) 936-1817}
        \HEAD{High Energy Physics}
        \PUBNUM{HE} \MICH@t{High Energy Physics}}
 \def\astro{\HEAD{Astrophysics}
            \PUBNUM{ASTRO} \MICH@t{Astrophysics}}
 \def\Lthree{\phonenumber={\rm (734) 764-4442}
        \faxnumber={(734) 936-6529}
        \HEAD{High Energy Physics}
        \PUBNUM{LEP3} \MICH@t{High Energy Physics}}
 \def\spin{\phonenumber={\rm (734) 936-1027}
        \faxnumber={(734) 936-0794}
	\HEAD{High Energy Spin Physics Group}
	\PUBNUM{SPIN} \MICH@t{High Energy Spin Physics Group}}
 \def\vax{\phonenumber={\rm (734) 936-2426}
	\HEAD{High Energy Physics Computing Services}
	\PUBNUM{VAX} \MICH@t{High Energy Physics Computing Services}}
 \def\ocs{\phonenumber={\rm (734) 764-3348}
	\HEAD{Office of Computing Services}
	\PUBNUM{OCS} \MICH@t{Office of Computing Services}}
 \def\RanBuild{\phonenumber={\rm (734) 764-4443}
        \faxnumber={(734) 936-1817}
        \HEAD{Randall Building Committee}
        \PUBNUM{HE} \MICH@t{Randall Building Committee}}
 \def\UM{\phonenumber={\rm (734) 764-4437}
        \HEAD{Department of Physics}
        \PUBNUM{PHYSICS} \MICH@t{Department of Physics}}
 
%%%
%%%           LETTER COMMAND MACROS
%%%
 
 \def\lettertext{\par\unvcopy\letterb@x\par}
 \def\multiletter{\setbox\letterb@x=\vbox\bgroup
       \everypar{\vrule height 1\baselineskip depth 0pt width 0pt }
       \singlespace \topskip=\baselineskip }
 \def\letterend{\par\egroup}
 
 \def\letters{\l@tt@rs{The University of Michigan}}
 \def\letter{\wlog{\string\letter}
        \vfill\supereject
        \normalspace       % set spacing for skipping
        \FRONTPAGE
        \nullbox{1pt}{1pt}{1pt}\vskip-0.8750in
        \setbox0 = \vbox{\hfuzz=50pt   % put the letterhead text in box0
           {\hsize=\headboxwidth
	   \unvcopy\headbox\hfill
           \singlespace        
           \vskip-0.1in
	   \dimen1=2.25truein   
	   \setbox1=\hbox{\hep(734) 763-4929}     \advance\dimen1 by \wd1
	   \setbox1=\hbox{\hep\the\phonenumber}   \advance\dimen1 by -1\wd1
           \centerline{\the\mylettername\hskip\dimen1{\hep\the\phonenumber}}}}
%
% This is an extract from the \insertUG macro of Dave Nitz.  (If it works,
% don't fix it.)  Put the seal in box1
%
 \ifse@l
   \dimen0=0.8in%                    seal box width (smaller than the real seal)
   \setbox1=\hbox to \dimen0{%
    \dimen0=1.0in%                   seal box height
    \vbox to \dimen0{
      \vss
%        \UGbody{tex$inputs:umseal.ps}{1.0}{8.31}{1}                 % vax
%        \UGbody{/usr/server/TeX/inputs/umseal.ps}{1.0}{8.31}{1}     % sun
        \UGbody{/usr/local/tex-0.3.3/texmf/tex/plain/misc/umseal.ps}{1.0}{8.31}{1}      % decs
%        \UGbody{/progs/tex/inputs/umseal.ps}{1.0}{8.31}{1}          % apollo
     }%
     \hss
    }
 \fi
        \hfuzz=5pt
        \ifse@l
          \hfuzz=15pt\line{\hbox to0pt{}\hskip 0.3in\box1\box0 \hfill}
	\else
	  \line{\hfill\box0\hfill}
	\fi
        \vskip0.35in
        \vskip\letterpush
        \rightline{\today}
        \addressee}

 \def\myletter{\l@tt@rs{\the\myletterheadline}\letter}
 
 \def\signed#1{\par \penalty 9000 \bigskip \dt@pfalse
   \everycr={\noalign{\ifdt@p\vskip\signatureskip\global\dt@pfalse\fi}}
   \setbox0=\vbox{\singlespace \halign{\tabskip 0pt \strut ##\hfill\cr
    \noalign{\global\dt@ptrue}#1\hfill\crcr}}
   \line{\hskip 0.5\hsize minus 0.5\hsize \box0\hfill} \medskip }
 \def\copies#1{\singlespace\hfill\break
   \line{\nullbox{0pt}{0pt}{\hsize}}
   \setbox0 = \vbox {
     \noindent{\tenrm cc:}\hfill %.2196
     \vskip-0.2115in\hskip0.000in\vbox{\advance\hsize by-\parindent
       \ialign to\hsize{\strut ##\hfill\hfill
                 \tabskip 0pt plus \hsize \cr #1\crcr}}
       \hbox spread\hsize{}\hfill\vfill}
   \line{\box0\hfill}}
 \def\endletter{\nullbox{0pt}{0pt}{\hsize}
       \ifnum\pagenumber=1
                \vskip\letterbottomfil\vfill\supereject
          \else
                \vfill\supereject
       \fi
       \wlog{ENDLETTER}}
 
%%%%
%%%% LETTER AND MEMO SLACISMS
%%%%

%%%%%
%%%%%              MEMO DEFINITIONS
%%%%%
 
 \def\SLACMEMO{\letterstyle\FRONTPAGE \letterfrontheadline={\hfil}
     \line{\quad\fourteenrm SLAC MEMORANDUM\hfil\twelverm\the\date\quad}
     \medskip \memod@f}
 \def\MEMO{\letterstyle\FRONTPAGE \letterfrontheadline={\hfil}
     \l@tt@rcheck
     \line{\fourteenrm \the\memoheadline \hfill
%UM \the\memolabel\ Memorandum\hfill
                \twelverm\the\date}
     \medskip \memod@f}
 
 \def\memit@m#1{\smallskip \hangafter=0 \hangindent=1in
       \Textindent{\caps #1}}
 \def\memod@f{\xdef\To{\memit@m{To:}}
              \xdef\from{\memit@m{From:}}%
              \xdef\topic{\memit@m{Topic:}}
              \xdef\subject{\memit@m{Subject:}}%
              \xdef\rule{\bigskip\hrule height 1pt\bigskip}}
 %%%%%%%%%%\memod@f
 
%%%%%
%%%%%               TELE-FAX MACROS
%%%%%
 
\def\FAX{\letterstyle\FRONTPAGE
    \letterfrontheadline={
          \seventeenrm\hfill\qquad OUTGOING FACSIMILE COVER SHEET\hfill}
    \line{\quad\fourteenrm University of Michigan: Physics\hfil
                                      Fax No. \the\faxnumber\fourteenrm\quad}
    \smallskip
    \line{\quad\fourteenrm Randall Lab\hfil\the\date\quad}
    \medskip \faxd@f}

\def\faxd@f{\xdef\numcall{\memit@m{At 'Fax:}}%
            \xdef\numpage{\memit@m{Pages:}}%
            \memod@f}
     %%%%%%%%%%%%\faxd@f
 
%%%%%
%%%%%             PAPER DEFINITIONS
%%%%%
 
%%%
%%%     PARAMETERS
%%%
 
 \newif\ifp@bblock          \p@bblocktrue
 \newif\ifpaperstyle
 \newif\ifb@@kstyle
 \newtoks\paperheadline
 \newtoks\bookheadline
 \newtoks\chapterheadline
 \newtoks\paperfootline
 \newtoks\bookfootline
 \newtoks\Pubnum            \Pubnum={$\caps UM - PHY - PUB -
                                        \the\year - \the\pubnum $}
 \newtoks\pubnum            \pubnum={00}
 \newtoks\pubtype           \pubtype={\tensl Preliminary Version}
 \newcount\yeartest
 \newcount\yearcount
 
%%%
%%%           STYLES
%%%
 \def\sequentialfootnotes{\global\seqf@@tstrue}
 \def\PHYSREV{\paperstyle\PhysRevtrue\PH@SR@V
     \let\refmark=\attach}
 \def\PH@SR@V{\doubl@true \baselineskip=24.1pt plus 0.2pt minus 0.1pt
              \parskip= 3pt plus 2pt minus 1pt }
 \def\IEEE{\paperstyle\IEEEtrue\I@EE\doublespace\rm\let\refmark=\IEEErefmark%
             \let\unnumberedchapters=\relax}
 \def\I@EE{\baselineskip=24.1pt plus 0.2pt minus 0.1pt
              \parskip= 3pt plus 2pt minus 1pt } 

%%%%%
%%%%%             ALLOW GREATER VARIATION OF STYLE DECLARATION
%%%%%

 \def\bookstyle{\b@@kstyletrue%
                \letterstylefalse%
                \paperstylefalse%
                \equ@tionlo@d
                \Tenpoint
                \frenchspacing
                \parskip=0pt
                \bookspace\booksize}
 \def\paperstyle{\paperstyletrue%
                 \b@@kstylefalse%
                 \letterstylefalse%
                 \normalspace}
%
%  the following changed on 4/27/88   B. Ball, J. Chapman
%
%                 \normalspace\papersize}
% \def\papersize{\hsize=35pc\vsize=50pc\hoffset=1pc\voffset=0.375in
%                \skip\footins=\bigskipamount}
 
 \def\booksize{\hsize=29pc\vsize=45pc\hoffset=0.85in\voffset=0.475in\hfuzz=2.5pc
               \itemsize=\parindent}
 
%%%
%%%           HEAD AND FOOT LINES
%%%
 
 \paperfootline={\hss\iffrontpage
                         \else \ifp@genum
                                  \hbox to \hsize{\tenrm\hfill\folio\hfill}\hss
                               \fi
                      \fi}
 \paperheadline={\hfil}
 \bookfootline={\hss\iffrontpage
                       \ifp@genum
                              \hbox to \hsize{\tenrm\bf\hfill\folio\hfill}\hss
                       \fi
                     \else
                           \hbox to \hsize{\hfill\hfill}\hss
                     \fi}
 
%%%%
%%%%                      TITLE PAGE MACROS
%%%%
 
 \def\titlepage{
    \yeartest=\year \advance\yeartest by -1900
    \ifnum\yeartest>\yearcount
          \global\PUBNUM{UM}
    \fi
    \FRONTPAGE\paperstyle\ifPhysRev\PH@SR@V\fi
    \ifp@bblock\p@bblock\fi}
 
 \def\nopubblock{\p@bblockfalse}
 \def\endpage{\vfil\break}
 \def\p@bblock{\begingroup \tabskip=\hsize minus \hsize
    \baselineskip=1.5\ht\strutbox \topspace-2\baselineskip
    \halign to\hsize{\strut ##\hfil\tabskip=0pt\crcr
    \the\Pubnum\cr \the\date\cr \the\pubtype\cr}\endgroup}
 
%%%
%%%      ACTIVE COMMANDS ON TITLE PAGE
%%%
 
 \def\PUBNUM#1{
     \yearcount=\year
     \advance\yearcount by -1900
     \Pubnum={$\caps UM- #1 - \the\yearcount - \the\pubnum $}}
 %%% DO NOT FORGET TO SET \pubnum={<number>}
 \def\title#1{\vskip\frontpageskip \titlestyle{#1} \vskip\headskip }
 \def\author#1{\vskip\frontpageskip\titlestyle{\twelvecp #1}\nobreak}

 \def\address#1{\par\noindent\titlestyle{\twelvepoint\it #1}}

 \def\andaddress{\par\kern 5pt \centerline{\sl and} \address}
 \def\abstract{\vskip\frontpageskip\centerline{\fourteenrm ABSTRACT}
               \vskip\headskip }

%%%%%
%%%%%               FOOTNOTE DEFINITIONS
%%%%%
 
%%%
%%%     PARAMETERS
%%%
 
 %%
 %%    parameters that are part of TeX
 %%        and don't need to be defined again
 %%
       %\newinsert\footins
       %\skip\footins=\bigskipamount % space added when footnote is present
       %\count\footins=1000 %footnote magnification factor (1 to 1)
       %\dimen\footins=8in % maximum footnotes per page
 %%
 %%
 
 \newtoks\foottokens       \foottokens={\Tenpoint\singlespace}
 \newdimen\footindent      \footindent=24pt
 \newcount\lastp@g@no	   \lastp@g@no=-1		 
 \newcount\footsymbolcount \footsymbolcount=0
 \newif\ifPhysRev
 \newif\ifIEEE
 \newif\ifseqf@@ts         \global\seqf@@tsfalse
 
%%%
%%%    CONTROL SEQUENCES
%%%
 
 %%
 %%       controls that are part of TeX
 %%           and don't need to be defined again
 %%
      %\def\fo@t{\ifcat\bgroup\noexpand\next \let\next\f@@t
      %       \else\let\next\f@t\fi \next}
      %\def\f@@t{\bgroup\aftergroup\@foot\let\next}
      %\def\f@t#1{#1\@foot}
      %\def\@foot{\strut\egroup}
      %\def\footstrut{\vbpx to\splittopskip{}}
 %%
 %%
 
 \def\footrule{\dimen@=\prevdepth\nointerlineskip
    \vbox to 0pt{\vskip -0.25\baselineskip \hrule width 0.35\hsize \vss}
    \prevdepth=\dimen@ }
 \def\vfootnote#1{\insert\footins\bgroup  \the\foottokens
    \interlinepenalty=\interfootnotelinepenalty \floatingpenalty=20000
    \splittopskip=\ht\strutbox \boxmaxdepth=\dp\strutbox
    \leftskip=\footindent \rightskip=\z@skip
    \parindent=0.5\footindent \parfillskip=0pt plus 1fil
    \spaceskip=\z@skip \xspaceskip=\z@skip
    \Textindent{$ #1 $}\footstrut\futurelet\next\fo@t}
 
%%%%
%%%%         ACTIVE USER COMMANDS
%%%%
 
 \def\footnote#1{\attach{#1}\vfootnote{#1}}
 
 \def\foot{\attach\footsymbolgen\vfootnote{^{\footsymbol}}}
 
%%%
%%%      FOOT SYMBOL CHOICES
%%%
 
 \let\footsymbol=\star
 \def\footsymbolgen
 {  \relax 
   \ifseqf@@ts \seqf@@tgen
   \else
      \ifPhysRev
         \iffrontpage \NPsymbolgen
         \else \PRsymbolgen
         \fi 
      \else \NPsymbolgen
      \fi
   \fi
   \footsymbol
 }
 \def\seqf@@tgen{\ifnum\footsymbolcount>0 \global\footsymbolcount=0\fi
       \global\advance\footsymbolcount by -1
       \xdef\footsymbol{\number-\footsymbolcount} }
 \def\NPsymbolgen
 {  \ifnum\footsymbolcount<0 \global\footsymbolcount=0\fi
   {  \iffrontpage \relax 
      \else
         \ifnum \lastp@g@no = \pageno
            \relax
         \else
            \global\lastp@g@no = \pageno
            \global\footsymbolcount=0 
         \fi
      \fi
   } 
   \ifcase\footsymbolcount
      \fd@f\star \or \fd@f\dagger \or \fd@f\ast \or
      \fd@f\ddagger \or \fd@f\natural \or \fd@f\diamond \or
      \fd@f\bullet \or \fd@f\nabla
   \fi
   \global\advance\footsymbolcount by 1
   \ifnum\footsymbolcount>6 \global\footsymbolcount=0\fi
 }
 \def\fd@f#1{\xdef\footsymbol{#1}}
 \def\PRsymbolgen{\ifnum\footsymbolcount>0 \global\footsymbolcount=0\fi
       \global\advance\footsymbolcount by -1
       \xdef\footsymbol{\sharp\number-\footsymbolcount} }
 \def\space@ver#1{\let\@sf=\empty \ifmmode #1\else \ifhmode
    \edef\@sf{\spacefactor=\the\spacefactor}\unskip${}#1$\relax\fi\fi}
 \def\attach#1{\space@ver{\strut^{\mkern 2mu #1} }\@sf\ }
 
%%%
%%%        HEAD AND FOOT LINE CHOICE
%%%
 
 \footline={\ifletterstyle \the\letterfootline
               \else \ifpaperstyle \the\paperfootline
                        \else \the\bookfootline
                      \fi
            \fi}
 \headline={\let\conbreak=\tableconbreakspace%
            \let\tbreak=\tableconbreakspace%
            \ifletterstyle \the\letterheadline
               \else \ifpaperstyle \the\paperheadline
                        \else \iffrontpage {}
                                  \else
                                     \setbox0=\hbox{\ {\tenrm\bf \folio}}
                                     \advance\hsize by \wd0
                                     \ifodd\pagenumber
                                          \hbox to \hsize{%
                                              \the\bookheadline\hfill\hfill
                                              \box0
                                                         }%
                                        \else
                                          \hskip-\wd0
                                          \hbox to \hsize{%
                                              \box0\hfill\hfill
                                              \the\chapterheadline
                                              }%
                                     \fi
                              \fi
                     \fi
             \fi}
 
%%%%%
%%%%%		Macros for making mailing labels.
%%%%%		#1 = vertical size of a single label; if empty, use 1.5in
%%%%%		#2 = number of labels across the page; if empty,  use 3
%%%%%		#3 = distance from left page edge to text start;
%%%%%		     if empty or 0in, then use a default starting 
%%%%%		     point for 1.5in x 3
%%%%%		#4 = distance from top of page to text top;
%%%%%		     if empty or 0in, then use a default starting 
%%%%%		     point for 1.5in x 3
%%%%%
%%%%%		A standard usage would be:
%%%%%		   \labelsetup{}{}{}{}
%%%%%		   \label{Dr. T.W. Jones\\Dept. of Physics \& Astron.\\
%%%%%		   University College London\\Gower St., London WC1\\
%%%%%		   UNITED KINGDOM }
%%%%%

\newdimen\vlsiz
\newdimen\hlsiz
\def\labelinit#1#2#3#4{
   \def\next{#1} \ifx\next\empty \vlsiz=1.5in \else \vlsiz=#1 \fi
   \hlsiz=\hsize
   \def\next{#2} \ifx\next\empty \divide\hlsiz by 3
                 \else           \divide\hlsiz by #2 \fi
   \advance\hlsiz by -0.01in
   \def\next{#3} \ifx\next\empty \else
   \ifdim#3=0pt \else \hoffset=-0.90in \advance\hoffset by #3 \fi \fi
   \def\next{#4} \ifx\next\empty \else
   \ifdim#4=0pt \else \voffset=-0.90in \advance\voffset by #4 \fi \fi}
\def\labelsetup#1#2#3#4{
   \nopagenumbers
   \singlespace
   \def\\{\ifnum\lastpenalty=1000 \relax\else\hfil\break\fi\ignorespaces}
   \hoffset=-.6in
   \voffset=-.5in
   \vsize=11in
   \hsize=8.5in
   \parskip=0pt
   \parindent=0pt
   \raggedbottom
   \def\label##1{\ifvmode\noindent\fi
      \vbox to \vlsiz{\hsize=\hlsiz\raggedright ##1\par\vss}
      \hskip 0pt plus.25in minus.25in\ignorespaces}
   \labelinit{#1}{#2}{#3}{#4}}

%%%%%%
%%%%%%          FIGURE INSERTION FOR QMS PRINTER
%%%%%%
 
%%%%     This macro command controls the insertion of figures
%%%%  into the TeX output for the Qms laser printer only!!!!
%%%%  The command asks for #1 the plotfiles name, #2 the
%%%%  verticle size of the plot, and #3 how far from the left
%%%%  margin the plot should be.
%%%%
%%%%    =====>> IMPORTANT: In order to have the plot printed,
%%%%               the plotfile must be in the same directory
%%%%               as the qms file and must be present while
%%%%               the dvi file is being processed.

%%  macro for inserting figures into the text
%%      #1   File where the plot information is located
%%      #2   Verticle size of the plot
%%      #3   How far from left margin to put plot
 
%%%%%%%
%%%%%%%              MASTER RESET FOR PHYZZEX MACRO
%%%%%%%
 
%%%%%                   SETS ALL GLOBAL DEFAULTS
 
 \def\masterreset{\global\pagenumber=1 \global\chapternumber=0
    \global\appendixnumber=0
    \global\equanumber=1 \global\sectionnumber=0 \global\subsectionnumber=0
    \global\referencecount=0 \global\figurecount=1 \global\tablecount=0
    \global\problemnumber=0
    \global\@ppendixfalse
    \ifIEEE\global\setbox\referencebox=\vbox{\normalbaselines 
          \noindent{\bf References}\vskip\headskip}%
     \else\global\setbox\referencebox=\vbox{\normalbaselines
          \centerline{\fourteenrm REFERENCES}\vskip\headskip}\fi
   }

%%%%%%
%%%%%%                 ANCIENT COMMANDS THAT ARE NO LONGER USED
%%%%%%
 
%%%
%%%           ACTIVE COMMAND
%%%
 
 \def\obsolete#1{\message{Macro \string #1 is obsolete.}}
 
%%%
%%%          OLD COMMANDS
%%%
 
 \def\product#1{\obsolete\product \prod_{#1}}
 \def\firstsec#1{\obsolete\firstsec \section{#1}}
 \def\firstsubsec#1{\obsolete\firstsubsec \subsection{#1}}
 \def\thispage#1{\obsolete\thispage \global\pagenumber=#1\frontpagefalse}
 \def\thischapter#1{\obsolete\thischapter \global\chapternumber=#1}
 \def\nextequation#1{\obsolete\nextequation \global\equanumber=#1
    \ifnum\the\equanumber>0 \global\advance\equanumber by 1 \fi}
 \def\BOXITEM{\afterassigment\B@XITEM\setbox0=}
 \def\B@XITEM{\par\hangindent\wd0 \noindent\box0 }
 
%%%%%%%%%%%%%%%%%%%%%%%%%%%%%%%%%%%%%%%%%%%%%%%
%%%%%%%%%%%%%%%%%%%%%%%%%%%%%%%%%%%%%%%%%%%%%%%
 
%%%%%
%%%%%             ALLOW THE USER TO INPUT OWN COMMANDS
%%%%%

%%%%%%%%%%%%%%%%%%%%%%%%%%%%%%%%%%%%%%%%%%%%%%%
%%%%%%%%%%%%%%%%%%%%%%%%%%%%%%%%%%%%%%%%%%%%%%%
%%%%%
%%%%%                 SET THE CHOOSEN LOCAL DEFAULTS
%%%%%
 
 \normalspace                      % set line spacing
 \masterreset
 \pagenumbers                      % default setting --> USE PAGE NUMBERS
 \Twelvepoint                      % default font size
 \figurelistoff                    % default for figure listing is off
 \tableconlistoff                  % default for contents listing is off
 \tablelistoff                     % default for table listing is off
 \reflist                          % default for ref listion is on
                                   %         for compatibility with PHYZZX
 \equationlistoff                  % default for equation listing is off
 \telex{4320815 UOFM UI}           % globally set the telex number
                                   % note: number must be set before the group
 \headboxwidth=6.5in
 \advance\headboxwidth by -1.05in  % Set a default letter-head width
 \UM                               % default publication type
 \paperstyle                       % This is the default
                                   % other styles: \letterstyle
                                   %               \bookstyle
 \tabledots                        % turn on dots for table of contents
                                   %   \notabledots turns them off
 \manualpageno                     % lists are automatically numbered
                                   %   \autopageno select starting page #
 \he@dboxsetfalse                  % do this to ensure proper date on letters
                                   %   and memos
 \lock
 \hfuzz=1pt
 \vfuzz=0.2pt
 
   % format identifiers
\message{PHYZZM version 1.9}

\PHYSREV

   %
% UGinsert parameters:
%	#1: name of UG generated postscript file
%	#2: figure width (inches)
%	#3: figure height (inches)
%	#4: left margin of figure (inches)
%	#5: bottom margin of figure (inches)
%	#6: desired figure magnification (1.0 = unmagnified)
%
% Parameters 2-5 refer to the dimensions when the UG generated
% PostScript file is printed as-is.
%
\def\UGinsert#1#2#3#4#5#6{
 \dimen0=#3in%
 \vbox to #6\dimen0{
 \vss
   \hbox to \hsize{%
     \hss
     \dimen0=#2in%
     \hbox to #6\dimen0{%
     \UGbody{#1}{#4}{#5}{#6}
     \hss
    }%
    \hss
   }
   \hss
  }
}
%
% The UGbody macro is included here for completeness, although it is now
% part of PHYZZM
%
% \def\UGbody#1#2#3#4{
%      \special {ps::[asis, begin]
%         0 SPB
%         save
%           /showpage {} def
%           /initgraphics {} def
%           Xpos Ypos translate
%           #4 dup scale
%           #2 72 mul neg #3 72 mul neg translate
%      }
%      \special{ps: plotfile #1 asis}
%      \special{ps::[asis,end]
%          restore
%          0 SPE
%      }
% }

   % Insert a UG figure with the figure caption on the right side of the
% figure.
%
% UGrcaption parameters:
%	#1: name of UG generated postscript file
%	#2: figure width (inches)
%	#3: figure height (inches)
%	#4: left margin of figure (inches)
%	#5: bottom margin of figure (inches)
%	#6: desired figure magnification (1.0 = unmagnified)
%	#7: figure caption
%
% Parameters 2-5 refer to the dimensions when the UG generated
% PostScript file is printed as-is.
%
\def\UGrcaption#1#2#3#4#5#6#7{
  \dimen0=#3in%
  \vbox to #6\dimen0{
    \dimen0=#2in
    \leftskip=#6\dimen0
    #7 \hfill
    \vss
    \UGbody{#1}{#4}{#5}{#6}
  }
}
%
% The UGbody macro is included here for completeness, although it is now
% part of PHYZZM
%
% \def\UGbody#1#2#3#4{
%      \special {ps::[asis, begin]
%         0 SPB
%         save
%           /showpage {} def
%           /initgraphics {} def
%           Xpos Ypos translate
%           #4 dup scale
%           #2 72 mul neg #3 72 mul neg translate
%      }
%      \special{ps: plotfile #1 asis}
%      \special{ps::[asis,end]
%          restore
%          0 SPE
%      }
% }

   % Version for dvi2ps driver
%
% Insert 2 UG drawings side by side
%
% UGtwoinsert parameters:
%	#1: name of left UG generated postscript file
%	#2: name of right UG generated postscript file
%	#3: figure width (inches)
%	#4: figure height (inches)
%	#5: left margin of figure (inches)
%	#6: bottom margin of figure (inches)
%	#7: desired figure magnification (1.0 = unmagnified)
%
% Parameters 3-6 refer to the dimensions when the UG generated
% PostScript file is printed as-is.  Both drawings must be the same
% size and have the same margins when printed as-is.
%
\def\UGtwoinsert#1#2#3#4#5#6#7{
  \dimen0=#4in%
  \vbox to #7\dimen0{
    \vss
    \hbox to \hsize{%
      \hss
      \dimen0=#3in%
      \hbox to #7\dimen0{%
      \UGbody{#1}{#5}{#6}{#7}
        \hss
      }%
      \hss
      \hbox to #7\dimen0{%
      \UGbody{#2}{#5}{#6}{#7}
        \hss
      }%
      \hss
    }
    \hss
  }
}
%
% The UGbody macro is included here for completeness, although it is now
% part of PHYZZM
%
% \def\UGbody#1#2#3#4{
%      \special {ps::[asis, begin]
%         0 SPB
%         save
%           /showpage {} def
%           /initgraphics {} def
%           Xpos Ypos translate
%           #4 dup scale
%           #2 72 mul neg #3 72 mul neg translate
%      }
%      \special{ps: plotfile #1 asis}
%      \special{ps::[asis,end]
%          restore
%          0 SPE
%      }
% }

   % Version for dvips driver
%
% Insert 2 UG drawings side by side
%
% UGtwocaption parameters:
%	#1: name of left UG generated postscript file
%	#2: name of right UG generated postscript file
%	#3: figure width (inches)
%	#4: figure height (inches)
%	#5: left margin of figure (inches)
%	#6: bottom margin of figure (inches)
%	#7: desired figure magnification (1.0 = unmagnified)
%	#8: left figure title
%	#9: right figure title
%
% Parameters 3-6 refer to the dimensions when the UG generated
% PostScript file is printed as-is.  Both drawings must be the same
% size and have the same margins when printed as-is.
%
\def\UGtwocapt#1#2#3#4#5#6#7#8#9{
  \dimen0=#4in%
  \vbox to #7\dimen0{
    \vss
    \hbox to \hsize{%
      \dimen0=#3in%
      \hbox to #7\dimen0{%
        \UGbody{#1}{#5}{#6}{#7}
        \hss
      }%
      \hss
      \hbox to #7\dimen0{%
        \UGbody{#2}{#5}{#6}{#7}
        \hss
      }%	
    }	
  }
  \smallskip
  \hbox to \hsize{%
    \dimen0=#3in%
    \setbox0=\hbox to #7\dimen0{\hsize=#7\dimen0\vtop{#8\hfill\vss}}%
    \setbox1=\hbox to #7\dimen0{\hsize=#7\dimen0\vtop{#9\hfill\vss}}%
    \box0 \hss \box1
  }
}
%
% The UGbody macro is included here for completeness, although it is now
% part of PHYZZM
%
% \def\UGbody#1#2#3#4{
%      \special {ps::[asis, begin]
%         0 SPB
%         save
%           /showpage {} def
%           /initgraphics {} def
%           Xpos Ypos translate
%           #4 dup scale
%           #2 72 mul neg #3 72 mul neg translate
%      }
%      \special{ps: plotfile #1 asis}
%      \special{ps::[asis,end]
%          restore
%          0 SPE
%      }
% }

\singlespace
\newbox\hdbox%
\newcount\hdrows%
\newcount\multispancount%
\newcount\ncase%
\newcount\ncols% This is the number of primary text columns in the table.
\newcount\nrows%
\newcount\nspan%
\newcount\ntemp%
\newdimen\hdsize%
\newdimen\newhdsize%
\newdimen\parasize%
\newdimen\thicksize%
\newdimen\thinsize%
\newif\ifendsize%
\newif\iffirstrow%
\newtoks\dbt%
\newtoks\hdtks%
\newtoks\savetks%
\newtoks\tableLETtokens%
\newtoks\tabletokens%
%
%  Book-keeping stuff--see how often these macros are called.
%
\immediate\write15{%
%CP SMSG GJMSINK TEXTABLE -----> TABLE MACROS LOADED, JOB = \jobname%
 -----> TABLE MACROS LOADED, JOB = \jobname%
}%
\catcode`\@=11%  Allows use of "@" in macro names, like PLAIN TEX does.
%  Debugging aid.  Writes #1 on the
%                                    user's terminal and in the log file.
\def\tstrut{\vrule height16pt depth6pt width0pt}%
\def\|{|}%  Make it easy to get |'s of type other after they are later
%           made active.
\def\tablerule{\noalign{\hrule height\thinsize depth0pt}}%
\thicksize=1.5pt%  Default thickness for fat rules.  The user should feel
%                  free to change this to his preference.
\thinsize=0.6pt%   Default thickness for thin rules.
\def\thickrule{\noalign{\hrule height\thicksize depth0pt}}%
\def\ctr#1{\hfil\ #1\ \hfil}%
%
%  \vctr vertically centers its argument in the row.
\def\tablewidth{}%
\parasize=4in%
\gdef\ARGS{########}%  Produces the correct number of #'s in the preamble
%                      by the time eveything is expanded and \halign sees
%                      it.
\gdef\headerARGS{####}%  Same as \ARGS, but used in \header macros.
\def\@mpersand{&}%  Allows us to get alignment tab characters later
%                   when we have made the character "&" an active macro.
{\catcode`\|=13%  Make |'s locally active.
\gdef\letbarzero{\let|0}%  Globally define a macro that allows us to
%                          keep active |'s from being expanded in edef's.
\gdef\letbartab{\def|{&&}}%  This \def will cause active |'s read by
%                            \ruledtable to be converted into double
%                            alignment tabs.
}%  End of locally active |'s.
{\def\ampskip{&\omit\hfil&}%  This local macro skips a vertical rule.
\catcode`\&=13%  Now make &'s into active macros.
\let&0%  This allows us to expand \ampskip in the next \xdef without
%        attempting to expand the & and getting an "undefined control
%        sequence" error.
\xdef\letampskip{\def&{\ampskip}}%  This will cause active &'s read by
%                                   \ruledtable to be converted into
%                                   double tabs and an \omit'ted \vrule.
}%  End of locally active &'s.
\def\begintable{%  Here we make |'s and &'s active characters so we can
%                  interpret them as macros.  Note that this action is
%                  true only until we encounter the matching \endgroup
%                  token later at the end of the \ruledtable macro.
   \begingroup%
   \catcode`\|=13\letbartab%
   \catcode`\&=13\letampskip%
   \def\multispan##1{%  We must redefine \multispan to count the number
%                       of primary columns, not physical columns.
      \omit \mscount##1%
      \multiply\mscount\tw@\advance\mscount\m@ne%
      \loop\ifnum\mscount>\@ne \sp@n\repeat%
   }%  End of \multispan macro.
   \def\|{%
      &\omit\widevline&%
   }%
   \ruledtable%  Now we call \ruledtable to do the real work.
}%  End of \begintable macro.
\newif\iflatetable \latetablefalse

\newbox\tablebox
% lets see if this works
\long\def\ruledtable#1\endtable{%
%
%  This macro reads in the user's data entries
%  and converts them into a ruled table.
%
%  Important note:  Many macros and parameters are re-defined here, and
%  these must be kept local to the table macros to avoid conflict with
%  their use outside of tables.  This is done by the \begingroup token
%  macro \begintable and the \endgroup token at the end of
%  this macro.
%
   \offinterlineskip%  Needed to make rules touch each other.
   \tabskip 0pt%  Needed for same reason as \offinterlineskip.
   \def\widevline{\vrule width\thicksize}%  Make outer \vrule's wider.
   \def\endrow{\@mpersand\omit\hfil\crnorm\@mpersand}%
   \def\crthick{\@mpersand\crnorm\thickrule\@mpersand}%
   \def\crnorule{\@mpersand\crnorm\@mpersand}%
   \let\nr=\crnorule%  A shorter abbreviation.
   \def\endtable{\@mpersand\crnorm\thickrule}%
   \let\crnorm=\cr%  Allows us to use \cr for our own purposes.
%
%  Cause user-typed \cr's to follow a row with a \tablerule.
%
   \edef\cr{\@mpersand\crnorm\tablerule\@mpersand}%
   \the\tableLETtokens%  Get the user's extra \let's, if any.
%
%  Put the data entries into a token register so we can scan through them
%  and see what the user is asking us to do.
%
   \tabletokens={&#1}%  We add an extra alignment tab to the beginning
%                       of the first row to allow for the first \vrule.
%
%  Now count how many rows are in the table and return the result in
%  count register \nrows; do the same for columns, and return that
%  in register \ncols.
%
   \countROWS\tabletokens\into\nrows%
   \countCOLS\tabletokens\into\ncols%
%
%  Now do a little arithmetic to convert the number of primary columns
%  into the number of physical columns that the alignment preamble must
%  prepare for;  similarly for rows.
%
   \advance\ncols by -1%
   \divide\ncols by 2%
   \advance\nrows by 1%
%
%  Tell the user how many rows and columns we found in his data.
%
   \immediate\write16{[Nrows=\the\nrows, Ncols=\the\ncols]}%
%
%  Now we actually go ahead and produce the table.
%
\iflatetable
   \global\setbox\tablebox=\hbox{%
	\tableframe%
	}% End of \hbox
\else
   \tableframe \fi
   \endgroup%  Return all local macros and parameters to their outside
%              values.
}%  End of macro \ruledtable.
\def\tableframe{%
   \line{%  The final table comes out as an \hbox of width the \hsize.
      \hss%  The final table will be centered left-to-right.
	\vbox{%
         \makePREAMBLE{\the\ncols}%  Generate the preamble.
         \edef\next{\preamble}%  This line and the next line force the
         \let\preamble=\next%    expansion of all \ARGS tokens into the
%                                appropriate number of #'s.
         \makeTABLE{\preamble}{\tabletokens}%  Go do the \halign here.
      }%  End of \vbox.
      \hss%  Finish the centering effect.
   }%  End of \line.
	}%	End of tableframe
\def\makeTABLE#1#2{%  Does an \halign for the \ruledtable macro.
   {%  Start of local parameter values.
   \let\ifmath0%     These macros would cause trouble if they were to be
   \let\header0%     expanded in the following \xdef; we \let them be
   \let\multispan0%  equal to a digit, because digits can't be expanded.
   \xdef\next{%  We must force the preamble to be expanded BEFORE the
      \halign\tablewidth{%
%        \halign is done;  this \edef\next{...}\next construction
%                does the trick.
      #1%  This is the preamble text.
      \noalign{\hrule height\thicksize depth0pt}%  Makes the top \hrule.
      \the#2\endtable%  This is the main body.
%
%     \noalign{\hrule height0.7pt depth0pt}%  Makes the last \hrule.
      }%  End of \halign.
   }%  End of \next.
   }%  End of local values.
   \next%  This \next must be outside of the local values, because now
%          we want those troublesome macros in the \let's above to have
%          their normal actions.
}%  End of macro \makeTABLE.
\def\makePREAMBLE#1{%  This macro generates the necessary preamble for a
%                      ruled table with #1 primary columns.
%                      (Primary columns means the number of columns NOT
%                       counting those used for vertical rules.)
   \ncols=#1%  Get the number of columns desired.
   \begingroup%  Start local parameter definitions.
   \let\ARGS=0%  This is the key to the whole thing; it prevents \ARGS
%                from being expanded in the followin \edef's.
   \edef\xtp{\widevline\ARGS&\tstrut\ctr{\ARGS}}%  A 1-column preamble.
   \advance\ncols by -1%  One column has been generated; decrement the
%                         counter.
   \loop%  Append as many further columns as needed to the preamble.
      \ifnum\ncols>0 %
      \advance\ncols by -1%
      \edef\xtp{\xtp&\vrule width\thinsize\ARGS&\ctr{\ARGS}}%
   \repeat
   \xdef\preamble{\xtp&\widevline\ARGS\crnorm}%  Adds the last \vrule.
   \endgroup%  End of local parameters.
}%  End of macro \makePREAMBLE.
\def\countROWS#1\into#2{%  This counts the number of rows in #1 by
%                          looking for control sequences that end a row,
%                          e.g., \cr, \crthick, etc., and puts the result
%                          into count register #2.
   \let\countREGISTER=#2%
   \countREGISTER=0%
%  \out{In countROWS:  tokens are [\the#1]}%
   \expandafter\ROWcount\the#1\endcount%
}%
\def\ROWcount{%
   \afterassignment\subROWcount\let\next= %
}%
\def\subROWcount{%
%  \out{In subROWcount:  next is [\meaning\next]}%  Debugging aid.
   \ifx\next\endcount %
      \let\next=\relax%
   \else%
      \ncase=0%
      \ifx\next\cr %
         \global\advance\countREGISTER by 1%
         \ncase=0%
      \fi%
      \ifx\next\endrow %
         \global\advance\countREGISTER by 1%
         \ncase=0%
      \fi%
      \ifx\next\crthick %
         \global\advance\countREGISTER by 1%
         \ncase=0%
      \fi%
      \ifx\next\crnorule %
         \global\advance\countREGISTER by 1%
         \ncase=0%
      \fi%
      \ifx\next\header %
%     \out{In subROWcount:  next=header, ncase set=1}%
         \ncase=1%
      \fi%
%     \out{In subROWcount:  ncase is [\the\ncase]}%
      \relax%
      \ifcase\ncase %
         \let\next\ROWcount%
%        \out{subROWcount---> ncase=\the\ncase}%
      \or %
         \let\next\argROWskip%
%        \out{subROWcount---> ncase=\the\ncase}%
      \else %
      \fi%
   \fi%
%  \out{subROWcount---> NEXT=\meaning\next}%
   \next%
}%  End of macro \subROWcount.
\def\counthdROWS#1\into#2{%
\dvr{10}%
   \let\countREGISTER=#2%
   \countREGISTER=0%
\dvr{11}%
%  \out{In counthdROWS:  tokens are [\the#1]}%
\dvr{13}%
   \expandafter\hdROWcount\the#1\endcount%
\dvr{12}%
}%
\def\hdROWcount{%
   \afterassignment\subhdROWcount\let\next= %
}%
\def\subhdROWcount{%
%\out{In subhdROWcount:  next is [\meaning\next]}%
   \ifx\next\endcount %
      \let\next=\relax%
   \else%
      \ncase=0%
      \ifx\next\cr %
         \global\advance\countREGISTER by 1%
         \ncase=0%
      \fi%
      \ifx\next\endrow %
         \global\advance\countREGISTER by 1%
         \ncase=0%
      \fi%
      \ifx\next\crthick %
         \global\advance\countREGISTER by 1%
         \ncase=0%
      \fi%
      \ifx\next\crnorule %
         \global\advance\countREGISTER by 1%
         \ncase=0%
      \fi%
      \ifx\next\header %
%\out{In subhdROWcount:  next=header, ncase set=1}%
         \ncase=1%
      \fi%
%\out{In subhdROWcount:  ncase is [\the\ncase]}%
\relax%
      \ifcase\ncase %
         \let\next\hdROWcount%
%\out{subhdROWcount---> ncase=\the\ncase}%
      \or%
         \let\next\arghdROWskip%
%\out{subhdROWcount---> ncase=\the\ncase}%
      \else %
      \fi%
   \fi%
%\out{subhdROWcount---> NEXT=\meaning\next}%
   \next%
}%
{\catcode`\|=13\letbartab
\gdef\countCOLS#1\into#2{%
%  \out{In countCOLS:  tokens are [\the#1]}
   \let\countREGISTER=#2%
   \global\countREGISTER=0%
   \global\multispancount=0%
   \global\firstrowtrue
   \expandafter\COLcount\the#1\endcount%
   \global\advance\countREGISTER by 3%
   \global\advance\countREGISTER by -\multispancount
%  \out{countCOLS-->[\the\countREGISTER]}
}%
\gdef\COLcount{%
   \afterassignment\subCOLcount\let\next= %
}%
{\catcode`\&=13%
\gdef\subCOLcount{%
%\out{In subCOLcount: next is [\meaning\next]}
   \ifx\next\endcount %
      \let\next=\relax%
   \else%
      \ncase=0
      \iffirstrow
         \ifx\next& %
            \global\advance\countREGISTER by 2%
            \ncase=0%
         \fi%
         \ifx\next\span %
            \global\advance\countREGISTER by 1%
            \ncase=0%
         \fi%
         \ifx\next| %
            \global\advance\countREGISTER by 2%
            \ncase=0%
         \fi
         \ifx\next\|
            \global\advance\countREGISTER by 2
            \ncase=0%
         \fi
         \ifx\next\multispan
            \ncase=1%
            \global\advance\multispancount by 1%
         \fi
         \ifx\next\header
            \ncase=2%
         \fi
         \ifx\next\cr       \global\firstrowfalse \fi
         \ifx\next\endrow   \global\firstrowfalse \fi
         \ifx\next\crthick  \global\firstrowfalse \fi
         \ifx\next\crnorule \global\firstrowfalse \fi
      \fi%  End of \iffirstrow.
\relax%\out{subCOL-->  ncase=[\the\ncase]}
% \out{subCOL-->  next=\meaning\next}
      \ifcase\ncase %
         \let\next\COLcount%
      \or %
         \let\next\spancount%
      \or %
         \let\next\argCOLskip%
      \else %
      \fi %
   \fi%
%  \out{subCOL-->  countREGISTER=[\the\countREGISTER]}
   \next%
}%
\gdef\argROWskip#1{%
%  Deletes the next balanced, undelimited argument from a
%                 token list.
% \out{---> Entering argROWskip <---}
% \out{In argROWskip:  deleted arg is [#1]}%
   \let\next\ROWcount \next%
}%  End of macro \argskip.
\gdef\arghdROWskip#1{%
%  Deletes the next balanced, undelimited argument from a
%                 token list.
% \out{---> Entering arghdROWskip <---}
% \out{In arghdROWskip:  deleted arg is [#1]}%
   \let\next\ROWcount \next%
}%  End of macro \arghdROWskip.
\gdef\argCOLskip#1{%
%  Deletes the next balanced, undelimited argument from a
%                 token list.
% \out{---> Entering argCOLskip <---}
% \out{In argCOLskip:  deleted arg is [#1]}%
   \let\next\COLcount \next%
}%  End of macro \argskip.
}%  End of active &'s.
}%  End of active |'s.
\def\spancount#1{%\out{spancount--->\meaning#1}
   \nspan=#1\multiply\nspan by 2\advance\nspan by -1%
   \global\advance \countREGISTER by \nspan
%  \out{number spancount--->\the\nspan; \the\countREGISTER}
   \let\next\COLcount \next}%
\def\dvr#1{\relax}%
% \omit\hfil%
% \parindent=0pt\hsize=1.1in\valign{%
% \vfil#\vfil&\vfil#\vfil\cr\hfil\hbox{\ Added to\ }\hfil&%
% \hfil\hbox{\ empty events\ }\hfil\cr}\hfil%
\def\header#1{%
\dvr{1}{\let\cr=\@mpersand%
\hdtks={#1}%
%\out{In header:  hdtks=[\the\hdtks]}%
\counthdROWS\hdtks\into\hdrows%
\advance\hdrows by 1%
\ifnum\hdrows=0 \hdrows=1 \fi%
%\out{In header:  Nhdrows=[\the\hdrows]}%
\dvr{5}\makehdPREAMBLE{\the\hdrows}%
%\out{In header:  headerpreamble=[\headerpreamble]}%
\dvr{6}\getHDdimen{#1}%
%\out{In header:  hdsize=[\the\hdsize]}%
%\striplastCR{#1}%
{\parindent=0pt\hsize=\hdsize{\let\ifmath0%
\xdef\next{\valign{\headerpreamble #1\crnorm}}}\dvr{7}\next\dvr{8}%
}%
}\dvr{2}}%  End of macro \header.
\def\makehdPREAMBLE#1{%This macro generates the necessary preamble for a
\dvr{3}%
%                      ruled table with \ncols primary columns.
%                      (Primary columns means the number of columns NOT
%                       counting those used for vertical rules.
\hdrows=#1%  Get the number of columns desired.
{%  Start local parameter definitions.
\let\headerARGS=0%
%  This is the key to the whole thing; it prevents \ARGS
\let\cr=\crnorm%
%                from being expanded in the followin \edef's.
\edef\xtp{\vfil\hfil\hbox{\headerARGS}\hfil\vfil}%
\advance\hdrows by -1%  One row has been generated; decrement the
%                         counter.
\loop%  Append as many further rows as needed to the preamble.
\ifnum\hdrows>0%
\advance\hdrows by -1%
\edef\xtp{\xtp&\vfil\hfil\hbox{\headerARGS}\hfil\vfil}%
\repeat%
\xdef\headerpreamble{\xtp\crcr}%
}%  End of local parameters.
\dvr{4}}%  End of \makehdPREAMBLE.
\def\getHDdimen#1{%
%\out{In getHDdimen:  Arg 1=[#1]}%
\hdsize=0pt%
\getsize#1\cr\end\cr%
}%  End of macro getHDdimen.
\def\getsize#1\cr{%
%\out{In getsize:  Arg 1=[#1]}%
%  Here we have to check arg#1 and see if the first token in #1 is an
%    \end; if so, we stop, else we check the width of arg#1.
%  We recall that each arg#1 will be terminated with a \cr token.
\endsizefalse\savetks={#1}%
%\out{In getsize:  the savetks = [\the\savetks]}%
\expandafter\lookend\the\savetks\cr%
%\out{In getsize:  ifendsize = [\meaning\ifendsize]}%
\relax \ifendsize \let\next\relax \else%
\setbox\hdbox=\hbox{#1}\newhdsize=1.0\wd\hdbox%
\ifdim\newhdsize>\hdsize \hdsize=\newhdsize \fi%
%\out{In getsize:  hdsize=[\the\hdsize]}%
%\out{In getsize:  newhdsize=[\the\newhdsize]}%
\let\next\getsize \fi%
\next%
}%
\def\lookend{\afterassignment\sublookend\let\looknext= }%
\def\sublookend{\relax%
%\out{In sublookend:  looknext = [\looknext]}%
\ifx\looknext\cr %
%\out{In sublooknext:  looknext=cr}%
\let\looknext\relax \else %
%\out{In sublooknext:  looknext/=cr}%
   \relax
   \ifx\looknext\end \global\endsizetrue \fi%
   \let\looknext=\lookend%
    \fi \looknext%
}%
%
%  Allow the user to make his own names for crthick, etc.
%
\def\tablelet#1{%
   \tableLETtokens=\expandafter{\the\tableLETtokens #1}%
}%
\catcode`\@=12%  Change @'s back to their normal category code.
\FRONTPAGE
\rightline{\the\date}
\medskip
   \title{Improved Probability Method for Estimating Signal in the
   Presence of Background}%
%\fi
   \author{Byron P. Roe$^1$ and Michael B. Woodroofe$^2$} %
\address{$^1$Department of Physics, University of Michigan, Ann Arbor,
MI 48109, \nextline $^2$Departments of  Mathematics and Statistics, 
University of Michigan, Ann Arbor, MI 48109} %
\medskip                                                           %
   \endpage                                                           %
\abstract
A suggestion is made for improving the  Feldman-Cousins\REF
\coustwo{G.J. Feldman and R.D. Cousins, Phys. Rev. {\bf D57}, 3873 (1998).}
\refend method of estimating signal counts in the presence of
background.  The method concentrates on finding essential
information about the signal and ignoring extraneous information about
background.  An appropriate method is found which uses the condition that 
that the number of background events obtained does not exceed the
total number of events obtained.  Several alternative approaches are
explored. 
\chapter{Introduction}
Feldman and Cousins\rlap,\refmark\coustwo
 in a recent article, have made major advances towards solving two 
long-standing problems concerning the use of
confidence levels for estimating a parameter from data.  
 The first of
 these is eliminating the bias 
that occurs when one  decides between using a confidence interval or a
 confidence bound, after examining 
the data.  The second is finding a confidence interval when the
 experimental result produces estimators 
that are close to or past known bounds for the parameters of interest.
 Feldman and Cousins' method 
is called the {\it unified approach} below and is described in Section 2.  
In the present paper we argue that the unified approach does not make quite
 enough of an allowance for the known 
bounds and suggest a modification.  The modification is illustrated with
 the KARMEN 2 Data\REF\karmen{B. Zeitnitz \etal, to be published in Prog. Part.
Nucl. Physics {\bf 40} (1997);\nextline
K. Eitel and B. Zeitnitz for the KARMEN collaboration, 
Proceedings Contribution to {\it Neutrinos '98}, Takayama, Japan, 
June 4-9, 1998.}\refend, where precisely this problem has arisen.
 The KARMEN group has 
been searching for a neutrino oscillation signal reported by an LSND
 experiment\rlap.\REF\lsndone{C. 
Athanassopoulos \etal, Phys. Rev. Lett. {\bf
75}, 2650 (1995);\nextline
C. Athanassopoulos \etal, Phys. Rev. Lett. {\bf 77}, 3082
(1996);\nextline
C. Athanassopoulos \etal, Phys. Rev. {\bf C54}, 2685
(1996);\nextline
W.C. Louis, representing LSND, Proceedings of the Erice School on
Nucl. Physics, 19th course, {\it Neutrinos in Astro, Particle and
Nuclear Physics}, 16--26 September, 1997;\nextline
Response to the FNAL PAC, the BooNE collaboration, April 15,
1998;\nextline
H. White, representing LSND, Neutrinos 98, Takayama, Japan, June 4-9
1998 {\it Results from LSND}.}\refend   
As of Summer 1998, they had expected to see $2.88\pm 0.13$ background 
events 
and $1.0$ - $1.5$ signal events, if the LSND results were real, but
 had seen no events.  From their 
analysis, they claimed to almost exclude the effect claimed by the
 LSND experiment.

	To be specific recall that the Poisson density with mean 
$\mu$ is
$$
	p_{\mu}(k) = {1\over k!}\mu^ke^{-\mu}\eqno{(1)}
$$
for $k = 0,1,2,\cdots$, and let $P_{\mu}$ denote the corresponding
distribution function, 
$P_{\mu}(k) = p_{\mu}(0)+\cdots+p_{\mu}(k)$.  Suppose that background
radiation is added to a signal 
producing a total observed count, $n$ say, that follows a Poisson
distribution with mean $b+\lambda$.  Here 
the background and signal are assumed to be independent Poisson random
variables, 
with means $b$ and $\lambda$ respectively.  What are appropriate
confidence intervals 
for $\lambda$ if no events are observed ($n = 0$) or, more generally,
if $n$ is smaller 
than $b$?  For $n = 0$ and a $90\%$ confidence level, the unified
intervals all have 
left endpoints at $\lambda = 0$, while the right endpoints decrease
from $2.44$ when 
$b = 0$ to $0.98$ when $b = 5$\rlap.\foot{We use here the published
numbers for $n=0$ given by Feldman and Cousins.  The numbers we obtain
differ slightly.  For $b=3,\ n=0$, we obtain 0.95 and for $b=5,\ n=0$,
we obtain 0.77.}  These are the right answers within
the formulation of the unified approach.  

	The formulation is suspect, however, because the confidence
intervals should not 
depend on $b$ when $n = 0$.  For if no events are observed, then both
the signal and background radiation must have been zero.  It is as if
two independent experiments were 
performed, one for the background and one for the signal.   
The fact that there were no background events may be interesting but it
is not 
directly relevant to inference about $\lambda$ once the signal is
known, and certainly the a priori expectation $b$ of
 the background radiation is irrelevant when one knows that 
the actual background was $0$.  In this case, the confidence interval 
for $\lambda$ should be the same as if one had observed a {\it signal}
of strength $0$--either $2.44$ using the unified approach, or $2.30$
using an upper confidence bound.  Statisticians have a name 
situations like this one.  The background radiation is called 
an {\it ancillary variable}, because its distribution does 
not depend on unknown parameters, and conventional statistical wisdom 
calls for conditioning on ancillary variables when 
possible\rlap.\REF\reid{Reid, N. {\it The roles
of conditioning in inference}, Stat. Sci. {\bf 10}, 
No. 2, 139-199 (1995).}\refend    
That is what we just did, since conditioning on no background events 
leaves $n$ as the signal.  

%Also, the modified intervals are not completely 
%independent of $b$ when $n = 0$, though they are far less sensitive 
%to $b$ than the unmodified intervals.  

	Our modification is described in Section 2, where it is 
compared to the unmodified procedure.  For the KARMEN 2 data the 
modified confidence region is substantially larger than the unmodified 
one and overlaps the major portion of the LSND region.
The modification is compared to a
 Bayesian solution in Section 4 and shown to agree with it quite well,
especially for low counts.  Some other possible modifications are discussed
briefly in Section 3.  Giunti\REF\giunti{C. Giunti, {\it Statistical
interpretations of the null result of the KARMEN 2 experiment}, DFTT
50/98 (hep-ph/9808405), August 25, 1998.}\refend has also proposed a
modification of the unified approach and applied it to the KARMEN 2
data.  Our approach is contrasted with his in Section 3.

\chapter{An Improved Method}

	It is not trivial to generalize the method just 
described to the case of non-zero counts $n$ that may be 
small compared to the expected background radiation.  
For if $n > 0$, then it is no longer possible to recover 
the background and signal.  The key to
 our modification is to remember that a confidence 
interval consists of values of the parameter that are 
consistent with the data (that is, are not rejected by an 
hypothesis test whose significance level is one minus the 
confidence level).  This is also the approach taken by Feldman and
Cousins.   Suppose, for example, that the expected backgound radiation
is $b = 3$ but that only one event is observed ($n = 1$).  
Is $\lambda = 2$ inconsistent with this observation?  From one 
point of view it is.  If $\lambda = 2$, then the probability of
observing at most one event is $e^{-5} + 5e^{-5} = 6e^{-5} = .040$, which 
is less than the usual levels of significance.  On the other hand, 
if only one event is observed, then there can have been at most one 
background event, and this information should be included 
in assessing significance.  For the probability of at most one 
background event, $e^{-3} + 3e^{-3} = 4e^{-3} = .199$, is not 
large, and if the statement $\lambda = 2$ is regarded as an 
hypothesis, then it seems unfair to include lower than expected 
background radiation as evidence against it.  The way to remove 
the effect of the low background radiation is to compute the 
conditional probability of at most one event (total), given at 
most one background event.  The latter is $6e^{-5}/4e^{-3} = 
1.5\times e^{-2} = .203$, which is not less than the usual 
levels of significance.  

	Some notation is required to adapt this reasoning 
to the unified approach.  The likelihood function 
in the signal plus background problem is $L_b(\lambda\vert n) = 
p_{b+\lambda}(n)$, where $n$ is the observed count.  Following 
Feldman and Cousins, let $\hat{\lambda} = \max[0,n-b]$ denote the 
maximum likelihood estimator of $\lambda$ and let
$$
	R_b(\lambda,n) = 
{L_b(\lambda\vert n)\over L_b(\hat{\lambda}\vert n)}\eqno{(2)}
$$
be the likelihood ratio statistic for testing $\lambda$.  Then the 
unified approach consists of taking those $\lambda$ for which 
$R(\lambda,n) \ge c(\lambda)$, where $c(\lambda)$ is the largest 
value of $c$ for which 
$$
	\sum_{k:R_b(\lambda,k) < c} p_{b+\lambda}(k) \le \alpha\eqno{(3)}
$$
and $1-\alpha$ is the desired confidence level.  In words, the 
left side of (3) is the probability that $R_b(\lambda,n) < c$; 
a level $\alpha$ generalized likelihood ratio test\REF\riceone{J.
Rice, {\it Mathematical Statistics and Data Analysis}, 2nd ed.,
Duxbury (1995).}\refend rejects the 
hypothesis $\lambda = \lambda_0$ if $R_b(\lambda_0,n) < c(\lambda_0)$;
and the unified confidence intervals consist of those $\lambda$ that 
are not rejected.  The modification suggested here consists of replacing 
$p_{b+\lambda}(k)$ by the conditional probability of exactly 
$k$ events total given at most $n$ background events.  
The latter is
$$
q_{b,\lambda}^n(k) = \cases{p_{b+\lambda}(k)/P_b(n)&if\ $k \le n$\cr
	\sum_{j=0}^{n} p_b(j)p_{\lambda}(k-j)/P_b(n)&if\ $k > n$,\cr}\eqno{(4)}
$$
since $k$ total events imply at most $n$ background events when 
$k \le n$.  Let $\tilde{R}^n_b(\lambda,k)$ denote the likelihood ratio
obtained using  
$q_{b,\lambda}^n(k)$; \ie,
$\tilde{R}^n_b(\lambda,k)=q^n_{b,\lambda}(k)/{\rm  
max}_{\lambda'}q^n_{b,\lambda'}(k)$.  Let $\tilde{c}_n(\lambda)$ be
the largest value of $c$ for which 
$$
	\sum_{k:\tilde{R}_b^n(\lambda,k) < c} 
q_{b,\lambda}^n(k) \le \alpha.\eqno{(5)}$$
Then the modified confidence interval consists of those 
$\lambda$ for which $\tilde{R}^n_b(\lambda,n) \ge \tilde{c}_n(\lambda)$.

	The modified and original unified approaches 
are compared in Figure 1 for the special 
case $b = 3$ and $n = 0,\cdots,15$.  Observe that the modified 
intervals are wider for small $n$ and that there is not much 
difference for large $n$.  The 
latter is to be expected, since there is not much difference 
between $q_{b,\lambda}^n$ and $p_{b+\lambda}$ for large $n$.  
In the case of small $n$, the rationale for the modification 
is as above.  If $n$ is smaller than $b$, then there was less 
background radiation than expected, and this information should 
be used in assessing significance.

\topinsert{
\vbox{
\bigskip
%\UGinsert{probadd2.ps}{8.0}{8.0}{0.4}{2.0}{0.4}
\UGinsert{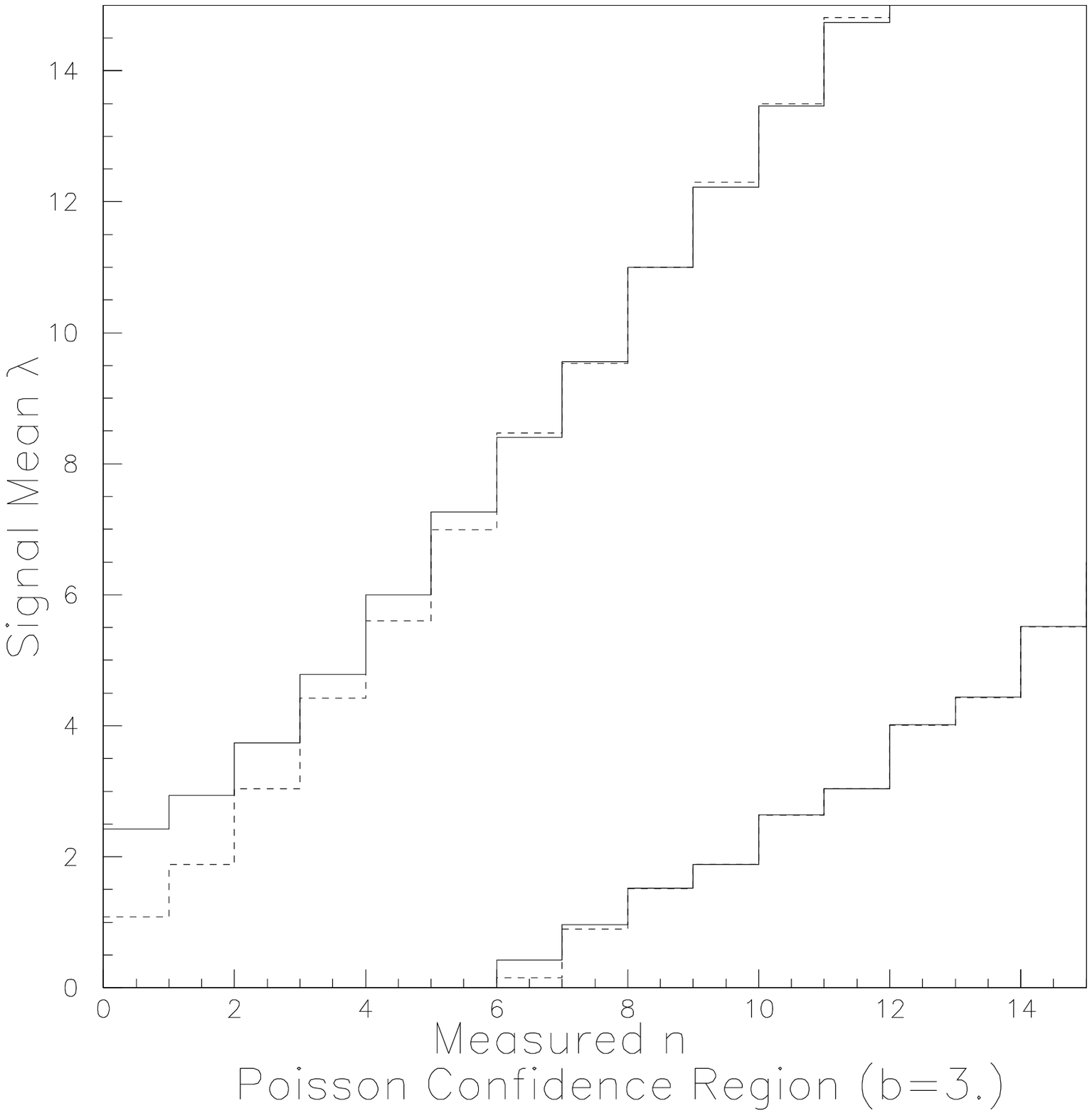}{8.0}{8.0}{0.5}{2.0}{0.5}
Figure 1.  The 90\% C.L. region for an unknown Poison signal 
$\lambda$ in 
the presence of a Poisson background $b=3$.  The dashed lines and solid
lines correspond to the unified approach and the modified 
approach, respectively.
}
\bigskip
}\endinsert

	For the KARMEN 2 Data, $b = 2.88\pm 0.13$ and $n = 0$.  
At the $90\%$ confidence level, the unified approach leads 
to $0 \le \lambda \le 1.08$, and the modified interval leads 
to $0 \le \lambda \le 2.42$.  As above, values of $\lambda$ 
between $1.08$ and
 $2.42$ are found to be inconsistent with the data 
by the unified approach, but this is due 
to lower than expected background radiation, and the 
inconsistency disappears after adjusting for the low 
background radiation.  On the basis of this data, it is not 
reasonable to exclude the possibility of signal.

	To be complete (and fair) Feldman and Cousins were aware of
the problem with small counts.  For such cases, they suggested
reporting the 
%confidence intervals that would have been used if $n$
%had been equal or close to $b$, 
the average upper limit that would be
obtained by an ensemble of experiments with the expected background and no
true signal, 
along with the intervals for the 
observed $n$.  The conceptual difference between our intervals and
the unified method is that our confidence levels are conditional and, 
therefore, refer to a different ensemble.  In general terms, the main 
reason for conditioning is to obtain a model that describes the 
experiment performed more accurately.  The price paid for the more 
accurate model is often a loss of power, or longer confidence
intervals, and the effect can be large, as in the KARMEN data.  
Of course, power is important, but it is an 
illusion if the model does not describe the experiment well.

	The reader may be familiar with conditioning in the context 
of contingency tables when some of the row and/or column totals are 
fixed, have known distributions, or have distributions that only 
depend on nuisance parameters.  In such cases it is appropriate to 
condition on the known totals, and this affects the distribution of 
tests statistics and estimators.  Fisher's exact test provides a 
specific example.  See Lehmann\REF\lehmann{E.L. Lehmann, 
{\it Testing Statistical Hypotheses}, 
2nd ed., pp. 156-162,  Wadsworth (1986).}
\refend for a derivation of 
the exact test and Berkson\REF\berkson{J. Berkson,   
{\it In dispraise of the exact 
test,} J. Statist. Plan. Inf. {\bf 2}, 27-40 (1978).}\refend 
for criticisms.  Other reasons for 
conditioning arise when the precision with which an experiment was 
done is observed as part of the outcome.  In the case $n = 0$, our 
use of conditioning is consistent with these precedents.  In the 
case $n > 0$, however, our use of conditioning goes beyond these 
established precedents because we condition on an observed bound 
for an ancillary variable, not the exact value.  Our reasons for 
conditioning, illustrated by the numerical example with $n = 1$ 
above, are consistent with the precedents.  To summarize these 
reasons: it seems unwise to regard lower than expected background 
radiation as evidence against a value of $\lambda$.

\chapter{Other Possible Modifications}
	The rationale given for the modification in Section 2 
could also have been used to support other modifications.  
We describe these briefly here and explain our preference for the 
one described in Section 2.  We also contrast our modification with
that of Giunti.

	The modification described in Section 2 replaces 
$p_{b+\lambda}$ with $q_{b,\lambda}^n$ in the {\it derivation} of the 
unified approach, thus replacing $R_b(\lambda,k)$ by 
$\tilde{R}_b^n(\lambda,k) = q_{b,\lambda}^n(k)/\max_{\lambda'}
q_{b,\lambda'}^n(k)$ in (2) and replacing Equation (3) by Equation (5).
An alternative modification would be to keep the 
unified approach criterion $R_b(\lambda,n)$ but calibrate 
the associated tests differently, by replacing $p_{b+\lambda}$ 
with $q_{b,\lambda}^n$ in Equation (3) and, therefore, $c(\lambda)$ 
with $c_n(\lambda)$ (except that $R$ not $\tilde{R}$ is used).  
We have explored this approach and found it to be 
very similar to the one 
presented.  It has the disadvantage that the limits for $n=0$ are
slightly dependent on $b$.  

Our approach may be contrasted with that of
Giunti\rlap,\refmark\giunti who has suggested a different modification
of the unified approach, called the {\it new ordering approach}.   His
physical arguments are along similar lines to ours.
However, in detail his approach differs.  In
the new ordering approach, $R_b(\lambda,n)$ is replaced by
$R^{NO}_b(\lambda,n)=p_{\lambda+b}(n)/p_{\lambda^{NO}+b}(n)$ in
Equation 3, where $\lambda^{NO}$ is the Bayes' estimate of $\lambda$
for a uniform prior.  (We shall describe the Bayes' approach further in
the next section.)  The calibration then proceeds as in Equation 3,
using $p_{b+\lambda}(k)$.  The resulting intervals are shorter than
ours, but depend on $b$ when $n=0$.  Amusingly, our intervals are
closer to the Bayesian intervals than are Giunti's intervals, even
though our approach is entirely frequentist.  See Table 1 below.

	In Equation (4), 
$$
	q_{b,\lambda}^n(n) = {p_{b+\lambda}(n)\over P_b(n)}\eqno{(6)}
$$
is the conditional probability of $n$ events (total) given at most $n$
background events.  This is a very intuitive quantity but,
unfortunately, is not a density in $n$, since $P_b(n) < 1$ for all $n$
and, therefore, $\sum_{n=0}^{\infty} q_{b,\lambda}^n(n
) > \sum_{n=0}^{\infty} p_{b+\lambda}(n) = 1$.  Of 
course, $q_{b,\lambda}^n(n)$ could be renormalized by 
$\kappa(\lambda) := \sum_{n=0}^{\infty} q_{b,\lambda}^n(n)$, 
and the resulting ratio $q_{b,\lambda}^n(n)/\kappa(\lambda)$ 
would be a density; but using the ratio in a model would implicitly 
change the likelihood function.  The density then lacks the intuitive
appeal of $q^n_{b,\lambda}(k)$, since the definition of the experiment
producing this density then becomes unclear.

	A closely related quantity is the conditional probability 
of at most $n$ events total, given at most $n$ background events
$$
	D_{b,\lambda}(n) = {P_{b+\lambda}(n)\over P_{b}(n)}.\eqno{(7)}
$$
It is not obvious, but $D_{b,\lambda}(n)$ is a distribution function 
in $n$ for reasons explained below.  Let 
$d_{b,\lambda}(n) = D_{b,\lambda}(n) - D_{b,\lambda}(n-1)$ denote the 
corresponding density.  Still another alternative is to replace 
$p_{b+\lambda}$ by $d_{b,\lambda}$ in the unified approach.  This too 
led to a procedure that was more complicated and no more efficient
than the modification described in Section 2.

	To see that $D_{b,\lambda}(n)$ is a distribution function in $n$, 
first observe that $\lim_{n\to\infty} D_{b,\lambda}(n) = 
\lim_{n\to\infty} P_{b+\lambda}(n)/P_{b}(n) = 1/1 = 1$.  So, it 
suffices to show that $D_{b,\lambda}(n)$ is non-decreasing in $n$.  
For this, note that, after some manipulation, $d_{b,\lambda}(n)$ can
be written in either of the following forms for $n>0$:
$$\eqalign{d_{b,\lambda}(n) &=d_{0b,\lambda}(n)-\coeff{\sum_{k=0}^{n-1}
p_{\lambda+b}(k)}
{\sum_{j=0}^{n-1}p_b(j)}\times\coeff{p_b(n)}{\sum_{i=0}^{n}p_b(i)}\cr
&=d_{0b,\lambda}(n)\left[1-\coeff{p_b(n)/\sum_{k=0}^{n-1}p_b(k)}
{p_{\lambda+b}(n)/\sum_{j=0}^{n-1}p_{\lambda+b}(j)}\right].\cr}\eqno{(8)},$$
where
$$d_{0b,\lambda}(n) ={p_{b+\lambda}(n) \over \sum_{j=0}^{n}p_b(j) }
\eqno{(9)}.$$
$D_{b,\lambda}(n)$ will be a non-decreasing function of $n$ if the
correction term in the second expression above is always $\le 1$.
Using the fact that these are Poisson distributions,
$$\eqalign{\coeff{p_b(n)/\sum_{k=0}^{n-1}p_b(k)}
{p_{\lambda+b}(n)/\sum_{j=0}^{n-1}p_{\lambda+b}(j)} &=
\coeff{e^{-b}b^n/n!} {\sum_{k=0}^{n-1}e^{-b}b^k/k! }\times
\coeff{\sum_{j=0}^{n-1}e^{-(b+\lambda)}(b+\lambda)^j/j!}
{e^{-(b+\lambda )}(b+\lambda)^n/n!}\cr
&=\coeff{\sum_{j=0}^{n-1}[1/(b+\lambda)^{n-j}j!]} 
{\sum_{k=0}^{n-1}[1/(b)^{n-k}k!]}\cr&\le 1.\cr}\eqno{(10)}$$
The last inequality occurs since $b+\lambda\ge b$.
        
\chapter{The Bayesian Connection}

	The discussion in this section makes use of the following 
identity, which may be established by repeated integrations by parts: 
if $m$ is any positive integer and $c \ge 0$, then
$$
	\int_c^\infty p_y(m)dy\equiv {1\over m!}\int_{c}^{\infty}
y^me^{-y}dy = 
\sum_{k=0}^m {1\over k!}c^ke^{-c}\equiv P_c(m),\eqno{(11)}
$$
This has an amusing consequence: While $q_{b,\lambda}^n(n)$ is 
not a density in $n$, it is a density in $\lambda$; that is,
$$
	\int_{0}^{\infty} q_{b,\lambda}^n(n)d\lambda = 1.\eqno{(12)}
$$
It follows that $q_{b,\lambda}(n)$ is the (formal) posterior 
distribution that is obtained when $\lambda$ is given an (improper) 
uniform distribution over the interval $0 \le \lambda < \infty$.  
(It is also the limiting posterior that is obtained if $\lambda$ is 
given a (proper) uniform distribution over the interval $0 
\le \lambda \le \Lambda$ and then $\Lambda$ is allowed to approach 
$\infty$).  Moreover, using (11) again, leads to the following curious 
relation
$$
	\int_{\lambda_0} q_{b,\lambda}^n(n)d\lambda =
	D_{b,\lambda_0}(n).\eqno{(13)}
$$
That is, the posterior probability that $\lambda$ exceeds $\lambda_0$ 
given $n$ is the conditional probability of at most $n$ events total 
given at most $n$ background events when $\lambda = \lambda_0$.  
Hence, using $D$, one of our possibilities above, although fully based
on a frequentist approach, has some Bayesian
justification.  Equation (13) also provides frequentist justification for 
conditioning.  For it follows from (13) and Theorem 3.3 of 
Hwang \etal\rlap,\REF\hwang{J.T. Hwang, G. Casella, C. Robert, M. Wells, and 
R. Farrell, {\it Estimation of accuracy in testing,}   
Ann. Statist., {\bf 20}, 490-509 (1992).}\refend 
that $D_{b,\lambda_0}(n)$ is an admissible $p$-value for testing 
$H_0: \lambda \ge \lambda_0$.  Admissibility of the unconditional 
$p$-value $P_{\lambda_0}(n)$ is unclear to us at this writing, 
if $b > 0$.

The Giunti approach, mentioned
above, fundamentally uses a partly Bayesian, partly frequentist approach.

	Treating $q_{b,\lambda}^n(n)$ as the posterior density in 
$\lambda$ leads to Bayesian credible (confidence) intervals of the 
form $\{\lambda: q_{b,\lambda}^n(n) \ge c_n\}$, where $c_n$ is so 
chosen to control the posterior probability of coverage; that is, 
$$
\int_{\{\lambda: q_{b,\lambda}^n(n) \ge c_n\}} 
q_{b,\lambda}^n(n)d\lambda = 1-\alpha.\eqno{(14)}
$$
Relation $(13)$ is useful in computing the latter integral.  The 
endpoints of these intervals have been computed for selected $b$ 
and $n$ and are compared to the endpoints of the modified unified 
approach in the table below.
\bigskip
Table 1.  Comparison of Confidence levels for the unified, modified
unified, Bayesian, and new ordering approaches described here, for $b=3$.
\medskip

%Using method 1
\begintable
 |\multispan{2}\tstrut\hfil
Unified\hfil|\multispan{2}\tstrut\hfil Modified\hfil|\multispan{2}\tstrut\hfil
Bayesian\hfil|\multispan{2}\tstrut\hfil
New Ord.\hfil\nr
$n$(observed) | Lower|Upper|Lower|Upper|Lower|Upper|Lower|Upper\crthick
0| 0.0  |1.08 |0.0  |2.42 |0.0  |2.30|0.0|1.86\nr
1| 0.0  |1.88 |0.0  |2.94 |0.0  |2.84|0.0|2.49\nr
2| 0.0  |3.04 |0.0  |3.74 |0.0  |3.52|0.0|3.60\nr
3| 0.0  |4.42 |0.0  |4.78 |0.0  |4.36|0.0|4.86\nr
4| 0.0  |5.60 |0.0  |6.00 |0.0  |5.34|0.0|5.80\nr
5| 0.0  |6.99 |0.0  |7.26 |0.0  |6.44|0.0|7.21\nr
6| 0.15|8.47 |0.42 |8.40 |0.0  |7.60|0.28|8.65\nr
7| 0.89|9.53 |0.96 |9.56 |0.55 |9.18|1.02|9.68\nr
8|1.51  |11.0 |1.52 |11.0 |1.20 |10.59|1.78|11.2\nr
9|1.88  |12.3 |1.88 |12.22|1.90 |11.91|2.49|12.4\nr
10|2.63 |13.5 |2.64 |13.46|2.63 |13.19|3.10|13.7\endtable
\medskip
\chapter{Summary}
We have suggested a modification to the unified approach of Feldman and
Cousins to further improve the estimation of 
signal counts in the presence of background.  It consists of replacing
the density function 
corresponding to the Poisson distribution $p_{b+\lambda}(k)$, with the
conditional density function $q^n_{b,\lambda}(k)$.  We noted that this
method has a clear frequentist justification and is the answer to a
clear statistics question.

We compared the results using this modification to the
unified approach with the results obtained using the unmodified
unified approach.
In contradistinction to the old method,
the new method leads naturally to sensible results if the observation
has fewer events than expected from background events alone.
\chapter{Acknowledgement}
We wish to thank Hsiuying Wang for bringing reference [\hwang] to our
attention. 
\endpage
\refout
\bye